
Figures in postscript form are available via anonymous FTP as follows.

ftp 130.113.0.111
anonymous
<email address>
cd pub
get fig1850.ps
quit

\magnification=\magstep1
\hsize = 6.5 true in\vsize = 9.0 true in
\baselineskip = 20 pt
\overfullrule=0pt
\def\titrule{\vrule height0.4pt depth0.0pt width4.0cm}
\footline={\hfil}
\headline={\ifnum\pageno=1 \hfil \else\hfil\folio \fi}

\newcount\tableone \tableone=0
\newcount\fignum \fignum=0
\newcount\fignumc \fignumc=0
\newcount\equa \equa=0
\newcount\itemm \itemm=0

\font\ftext=cmr10
\font\fsmall=cmr7

\def\taone{\global\advance\tableone by 1 \the\tableone}
\def\taonen{\the\tableone{\ (cont.)}}
\def\itemno{\global\advance\itemno by 1 \the\itemoone}
\def\takeep{\the\tableone}
\def\fig{\global\advance\fignum by 1 \the\fignum}
\def\figcc{\global\advance\fignumc by 1 \the\fignumc}
\def\equat{\global\advance\equa by 1 \the\equa}
\def\itemno{\global\advance\itemm by 1 \the\itemm}
\def\itemn{\item{\itemno)}}

\newdimen\digitwidth\setbox0=\hbox{\rm0}\digitwidth=\wd0
\catcode`~=\active\def~{\kern\digitwidth}
\catcode`!=\active\def!{\kern 0.7\digitwidth}
\catcode`|=\active\def|{\kern 1.4\digitwidth}

\edef\innernewbox{\noexpand\newbox}
\def\sbl{
\innernewbox\tablebox
\centerline { \box\tablebox}
\setbox\tablebox = \vbox \bgroup\offinterlineskip
\halign}

\def\tablerule{\noalign{\hrule}}
\def\pmm#1#2{|^{#1}_{#2}}

\def\deg{${}^\circ$}
\def\min{${}^{\prime}${ }}
\def\sec{${}^{\prime\prime}${ }}
\def\sol{$_\odot$}
\def\et{{\it et al.\ }}
\def\Rc2{{\it R}}
\def\vave{\overline {\hbox{v}}}

\def\kms{km s$^{-1}$}

\def\tabhl#1#2{{\ftext
\centerline{Table #1}
\centerline{#2}}
\innernewbox\tablebox
\centerline { \box\tablebox}
\setbox\tablebox = \vbox \bgroup\offinterlineskip
\halign\bgroup}

\edef\inskip{\noexpand\newskip}

\long\def\tabh#1#2{{\ftext
\centerline{Table #1}
\centerline{#2}}
\innernewbox\tablebox
\setbox\tablebox = \vbox \bgroup
\halign\bgroup}

\def\sp{\noalign{\vskip.2truecm}
\tablerule
\noalign{\vskip0.1truecm}
\tablerule\egroup\egroup
\centerline{\box\tablebox}}
\def\spc{\noalign{\vskip.2truecm}
\tablerule\egroup\egroup
\centerline{\box\tablebox}}

\def\spl{
\noalign{}
\tablerule
\noalign{\vskip0.1truecm}
\tablerule\egroup\egroup
\centerline {\box\tablebox}}

\def\splc{
\noalign{}
\tablerule\egroup\egroup
\centerline {\box\tablebox}}

\def\topper{
\noalign{\vskip0.3truecm}
\tablerule
\noalign{\vskip.1truecm}
\tablerule
\noalign{\vskip.1truecm}
}
\def\topperc{
\noalign{\vskip0.3truecm}
\tablerule
\noalign{\vskip.1truecm}
}
\def\topperl{
\noalign{\vskip0.3truecm}
\tablerule
\noalign{\vskip.1truecm}
\tablerule
}
\def\spacer{
\noalign{\vskip0.2truecm}
\tablerule
\noalign{\vskip0.1truecm}
}

\xdef\amper{&}
\newcount\n
\def\hspa#1{}
\def\tcoli{## \hfil}
\def\tcol#1{\hskip#1cm\hfil ## \hfil}
\def\tcolil#1#2{## \hfil\vrule height #1pt depth #2pt}
\def\tcoll#1#2#3{\hskip#1cm\hfil ## \hfil\vrule height #2pt depth #3pt}
\def\vbs#1#2{\hfil\vrule height #1pt depth #2pt}

\def\mcol#1#2{\n=0
\loop\ifnum\n<#2 \advance\n by1
\hskip#1cm\hfil ## \hfil \noexpand\amper \repeat}
\def\figc{\noindent Fig. \figcc --\ }
\pageno=1
\pageinsert
\baselineskip = 15pt
\centerline{\bf DYNAMICS OF THE}
\centerline{\bf YOUNG BINARY LMC CLUSTER NGC 1850}
\bigskip\bigskip
\centerline{PHILIPPE FISCHER$^{1}$, DOUGLAS L. WELCH$^{1}$}
\centerline{Department of Physics and Astronomy, McMaster University, Hamilton,
Ontario L8S
4M1, Canada}
\centerline{Email: Fischer@Crocus.Physics.McMaster.CA}
\centerline{ }
\centerline{MARIO~MATEO$^{2}$}
\centerline{The Observatories of the Carnegie Institute of Washington}
\centerline{813 Santa Barbara Street, Pasadena, CA 91101}
\vskip 0.50 truein
\newdimen\addwidth\setbox0=\hbox{Address for proofs: }\addwidth=\wd0
\vskip 0.25 truein
\leftline{Address for proofs:}
\leftline{Philippe Fischer}
\leftline{Department of Physics and Astronomy}
\leftline{McMaster University}
\leftline{1280 Main Street West}
\leftline{Hamilton, Ontario}
\leftline{L8S 4M1  CANADA}
\leftline{Email: Fischer@Crocus.Physics.McMaster.CA}
\vfil
\line{ \titrule \hfill}
\noindent
$^1$ Guest Investigator, Mount Wilson and Las Campanas Observatories,
which are operated by the Carnegie Institution of Washington.
\centerline{ }
\noindent
$^2$ Hubble Fellow.
\baselineskip = 20 pt
\endinsert
\centerline{ABSTRACT}
\medskip

In this paper we have examined the age and internal dynamics of the young
binary LMC cluster NGC 1850 using BV CCD images and echelle spectra of 52
supergiants.

Isochrone fits to a BV color-magnitude diagram revealed that the primary
cluster has an age of $\tau = 90 \pm 30$ Myr while the secondary member has
$\tau = 6 \pm 5$ Myr. The reddening was found to be E(B-V)=0.17 mag. BV
surface brightness profiles were constructed out to R $>$ 40 pc, and
single-component King-Michie (KM) models were applied.  The total cluster
luminosity varied from L$_B$ = 2.60 - 2.65 $\times 10^6$ L$_B$\sol\ and
L$_V$ = 1.25 - 1.35 $\times 10^6$ as the anisotropy radius varied from
infinity to three times the scale radius with the isotropic models providing
the best agreement with the data. Simple tests were made to check for tidal
truncation in the profiles and we concluded that there was slight evidence
favoring truncation. The bright background and binary nature of NGC 1850
render this conclusion somewhat uncertain.

Of the 52 stars with echelle spectra, a subset of 36 were used to study the
cluster dynamics. The KM radial velocity distributions were fitted to these
velocities yielding total cluster masses of 5.4 - 5.9 $\pm 2.4 \times 10^4$
M\sol\ corresponding to M/L$_B$ = 0.02 $\pm 0.01$ M\sol/L$_B$\sol\ or
M/L$_V$ = 0.05 $\pm 0.02$ M\sol/L$_V$\sol.  A rotational signal in the radial
velocities has been detected at the 93\% confidence level implying a
rotation axis at a position angle of 100\deg. A variety of rotating models
were fit to the velocity data assuming cluster ellipticities of $\epsilon =
0.1 - 0.3$.  These models provided slightly better agreement with the radial
velocity data than the KM models and had masses that were systematically
lower by a few percent.

Values for the slope of the mass function were determined using the
derived M/L, theoretical mass-luminosity relationships, and several forms
for the IMF. The preferred value for the slope of a power-law IMF is a
relatively shallow, $x = 0.29 \pmm{+0.3}{-0.8}$ assuming the B-band M/L or
$x = 0.71 \pmm{+0.2}{-0.4}$ for the V-band.
\medskip
\centerline{1. INTRODUCTION}
\medskip

There is a large population of young, massive clusters in the LMC that have
no counterparts in the Milky Way. These objects represent a unique
opportunity to study the internal dynamics of resolved stellar systems in
which the current ages are substantially less than the two-body relaxation
timescales. The clusters can be studied kinematically to determine masses
and mass-to-light ratios (hereafter, M/L's) in order to constrain the
initial mass function. Through the use of both imaging and spectroscopic
data one can determine the value of various dynamical parameters and better
understand formation and early cluster evolution.

To date there have been a number of dynamical studies of LMC clusters
including integrated spectra for several old and intermediate age clusters
(Elson and Freeman 1985; Dubath \et 1990; Mateo \et 1991; and Meylan \et
1991), individual stellar radial velocity measurements of intermediate age
clusters (Seitzer 1991, Fischer \et 1992b) and individual velocity
measurements for young clusters (Lupton \et 1989, Mateo \et 1991, Fischer
\et 1992a). Assuming stars can be resolved right to the cluster core, the
individual velocity measurements are favored over the integrated spectra.
With stellar radial velocities one can include radial information in the
dynamical models, rotation can be detected and quantified, escaping stars
can be detected, and, if multi-epoch observations are available, favorably
aligned binary stars can be found.

NGC 1850 is a very bright young cluster in the Bar region of the LMC. It is
located in a region rich with star clusters and star formation and is, in
fact, embedded in the emission nebula Henize 103. Furthermore, the cluster
appears to be in a binary or perhaps even a triple system, and there is a
clear indication of tidal interactions. The time is ripe for both photometric
and kinematic studies of this object. Only recently have complete stellar
isochrones appeared for high mass stars enabling one to determine ages more
reliably than in the past. The use of photon-counting detectors enables
high-precision radial velocity measurements for stars with apparent
magnitudes as faint as V = 18 mag.

In \S 2.1 the CCD imaging is described and, in \S 2.2, a color-magnitude
diagram is constructed and analyzed in order to derive an accurate age
estimate. In \S 2.3 surface density profiles are derived and in \S 3
King-Michie models are applied to this data. \S 4 contains a description of
the radial velocity observations and reductions. \S 5 has the cluster mass
estimates and \S 6 is a discussion of the cluster mass function. Finally \S
7 contains a calculation of the relaxation timescales for the cluster.
\medskip
\centerline{2. CCD IMAGING}
\medskip

\centerline{\it 2.1 The Data}
\medskip

BV CCD frames of NGC 1850 were obtained at the Las Campanas Observatory
(LCO) 1.0 m telescope on 1991 February 23.  The TEK2 1024$^2$ chip was used
(readout noise = 7 e$^-$, gain = 2 e$^-$/ADU, and angular scale =
0.61$^{\prime\prime}$ px$^{-1}$). The exposure times were 60 seconds for
each filter.

There are many interesting objects in the immediate vicinity (i.e. within
about 6 arcminutes) of NGC 1850 as can be seen from Figs. 1 and 2. Fig. 1
displays intensities around the median for the frame while Fig. 2 shows only
the brighter stars.  In Fig. 1 the cluster appears highly asymmetric with a
large collection of stars slightly north of due east. This has been
interpreted as both an interaction tail due to the presence of a binary
companion (Bhatia and MacGillivray 1987) and as a cluster with a population
distinct from the central cluster H88-159 (Bica \et 1992, based on
integrated photometry).  Following the line joining the main cluster to this
subcluster about 200 arcsec one finds a very low luminosity cluster (marked
with a 1 on the figure). Approximately 100 arcsec south there is a diffuse
object (2). Another, slightly higher luminosity cluster is located about 275
arcsec north of NGC 1850, and NGC 1855/54 (3), a bright young cluster, lies
to the south, with just its edge visible in this frame (4). Another
interesting feature is revealed in Fig. 2 directly to the West of NGC 1850
with a center-to-center separation of 30 arcsec (circled).  This object has
been previously designated by Bhatia \& Hatzidimitriou (1988) as a distinct
binary companion to NGC 1850 and according to Bica \et is the probable
source of ionizing radiation for the emission nebula Henize 103 in which
these objects are embedded. Bhatia \& Hatzidimitriou also concluded that the
frequency with which apparent double clusters occured in the LMC was much
higher than one would expect from chance superpositions.

\medskip
\centerline{\it 2.2 The Color-Magnitude Diagram}
\medskip

Stellar photometry was performed using DoPHOT (Mateo \& Schechter 1989). As
no standard stars were observed on this night the calibration was
accomplished through the use of another set of NGC 1850 frames taken on 1990
November 5 using the same telescope, CCD and filters as above. On that night
12 BV observations of 9 E-region standards (Menzies \et 1989) were observed
covering an airmass range of 1.02 - 1.52 and a color range of -0.043 mag
$\le$ B-V $\le$ 1.890 mag. The rms of the adopted solutions was approximately
0.02 mag.  Aperture corrections yielded a zero-point with a similar accuracy
and we conservatively estimate our absolute calibration to be accurate at
better than 0.04 mag. Stars were then matched up between the two sets of
frames to calibrate the present data. The reason for not using the older
data for the current work is that the more recent frames had superior seeing
which yielded improved stellar photometry particularly in the cluster's
inner regions.

There have been at least three previous published photometric studies of NGC
1850.  Robertson (1974) presented (BV)$_{pg}$ photographic photometry using
an iris photometer and based his calibration on the (BV)$_{pe}$ photoelectric
photometry of Tifft and Snell (1971). Alcaino \& Liller (1987) presented
photographic BVRI photometry also using an iris photometer and based their
calibration on a 15 star photoelectric sequence (Alcaino \& Liller 1982).
Elson (1991) performed BV CCD photometry and based her calibration on four
stars from Graham (1982), claiming a zero-point uncertainty of less than
0.01 mag. We have plotted the differences between our photometry and 44 stars
from Robertson, 99 stars from Alcaino \& Liller, and 190 stars from Elson in
Figs. 3 through 5, respectively. The mean zero-point differences, in the
sense of us--them, are $\Delta$V$_{Robertson} = 0.29$,
$\Delta$B$_{Robertson} = 0.36$, $\Delta$V$_{A \& L} = 0.06$, $\Delta$B$_{A
\& L} = 0.03$, and $\Delta$V$_{Elson} = -0.14$, $\Delta$B$_{Elson} = -0.10$,
We are fainter than Robertson and Alcaino \& Liller, brighter than Elson,
redder than both Robertson and Elson and bluer than Alcaino \& Liller.
There is a strong trend in both the B and V magnitude differences between
ourselves and Robertson in the sense that the agreement becomes
systematically worse for the fainter objects. Since this trend is not
significantly present for the other two comparisons we conclude that the
Robertson data is incorrect.

Fig. 6 shows an apparent BV color-magnitude diagram (CMD) for the entire
10\min square field. There is evidence for at least three distinct
populations.  Region 1 contains only very young stars still on the main
sequence. Region 2 contains a mixture of the very young stars as well as a
slightly older population of main sequence and evolved stars. Region three
appears to consist of a significantly older population of evolved stars.

Fig. 7 contains a small subset of the stars found in Fig. 6. These are all
the stars which were detected within 12 arcsec of the center of the binary
companion located 30 arcsec to the west of the main cluster. In fact, there
are many more stars in this region but unfortunately the inner region of the
companion cluster was saturated on the 60s frames. Even with a shorter
exposure we suspect that the extremely high degree of crowding in this
region would render stellar photometry both difficult and unreliable. In any
case, one can see that nearly all the detected stars are blue and they extend
beyond the clearly delineated turnoff region obvious in Fig. 6 indicating a
very young system.

We attempted to fit isochrones from Mermilliod \et 1992 to the photometry
data. These isochrones incorporate the new radiative opacities from Rogers
\& Iglesias (1992), convective overshooting and mass-loss (see Schaller
\et 1992 for a complete description of the stellar models). Isochrones were
available for z=0.0188 (solar abundance), z = 0.008, and z=0.001. There are
two published photometric metallicity determinations for NGC 1850:  the
first based on Washington photometry of two stars yielding z = 0.009 (Schommer
\& Geisler 1986), and the second based on Str\"omgren photometry of five
stars yielding z = 0.004 (Grebel \et 1992).

In Fig. 7 one can see the result of the isochrone fitting to the stars of
the companion cluster. The inferred reddening was E(B--V) = 0.17 $\pm 0.03$
mag [A$_V = 3.1$E(B--V) was assumed]. The presence of the young,
barely-evolved, companion provides an excellent opportunity to accurately
determine the reddening and our value agrees well with three previous
reddening determinations. Persson \et (1983) find E(B--V) = 0.15 mag using
the technique outlined in Cohen \et (1981) involving both optical and IR
broadband color indices. Alcaino and Liller (1987) obtained E(B--V) = 0.18
mag from a BVRI CMD study and Lee (1991) found E(B--V) = 0.15 $\pm$ 0.05 for
UBV photometry. The best age is $\tau = 6 \pm 5$ Myr. This changes by less
than 1 Myr as one varies the distance modulus by $\pm 0.3$ mag.

Fig. 8 is a CMD for the region within 1\min of the center of the main
cluster (it includes all the stars in Fig. 7). The apparent BV magnitudes
for this inner region are presented in Table 1.  Column 1, 6 and 11 are the
stellar identifications, column 2, 7 and 12 are the X positions on the CCD
(increases towards the south), column 3, 8 and 13 are the Y positions
(increasing towards the east), columns 4, 9, and 14 are the apparent V
magnitudes, and columns 5, 10, and 15 are the apparent B-V magnitudes.

The two younger isochrones in Fig. 8 are identical to those discussed above
while the older set are for $\tau = 90$ Myr. We have adopted the reddening
obtained above using the young stars.  Our age estimate is therefore $\tau =
90 \pm 30$ Myr, where the error estimate takes into account the uncertainty
in the zero-point and reddening and the difficulty in fitting the isochrones
to this crowded data set.  Varying the distance modulus by $\pm 0.3$ affects
the age estimate by approximately 10 Myr. Fig 9 shows population-synthesis
Monte Carlo simulations based on the two different isochrones with an
assumed photometric scatter of 0.05 mag and a mass function slope of 1.0 (see
equation 15). One feature is immediately apparent in both cases, the lack of
stars in the Hertzsprung gap compared to the observed data. Accepting the
correctness of the models, there are at least three possible explanations
for this effect:  1) The presence of young LMC field stars (and possibly a
few stars from the young subcluster), 2) Binaries and close superpositions
that were measured as single stars and 3) A possible time span for the
initial star formation (this would need to be at least half the cluster
age). We feel the first two effects are the likeliest and a close
superposition of stars has been invoked by Welch \& Stetson (1993) to
explain the anomalously small luminosity amplitude and high brightness of
three cepheids in the crowded regions of the young LMC cluster NGC 1866.
Another feature is that the evolution slows down around the red giant branch
(RGB) and at the blue end of the horizontal branch (HB) causing stars to
build up in these regions. This is observed and we conclude that the stars
are more metal-poor than z = 0.008 based on a synthetic HB which does not
extend far enough to the blue and a synthetic RGB which appears to be too
red. The z = 0.001 isochrone appears to have too blue a RGB although it does
provide a reasonable fit at the blue end of HB.  Therefore, the metallicity
lies between z=0.008 and z=0.001 but this conclusion is dependent on the
adopted reddening and distance modulus.

Previous age estimates are $\tau = 40 \pm 10$ Myr (Hodge 1983, based on data
from Tifft \& Connoly 1973, and Robertson 1974), $\tau = 21 \pm 5$ Myr
(Alcaino \& Liller 1987) and $\tau = 40 \pmm{+50}{-10}$ (Lee 1991). We have
shown that the the Alcaino \& Liller photometry and reddening estimate are
both close to our own and therefore the different age estimates can be
attributed to the different isochrones (they used the isochrones of Maeder
\& Mermilliod 1981) and the different age-determination techniques.
Unfortunately, the photometry of Lee was not available for comparison.

Finally, we examined stars in the region of H88-159. Unfortunately, there are
no stars in this region as bright as the turn-off for the main cluster.
Therefore, it is impossible to derive an accurate age for this subcluster in
order to determine if it is, in fact, a distinct population from the main
cluster as indicated from the integrated photometry of Bica \et (1992). An
argument against it being distinct as opposed to being a tidal tail is that
it appears to continue outward from the main cluster in a fairly straight
line for about 2 arcmin (see Fig. 1).

\medskip
\centerline{2.3 Surface Photometry}
\medskip

Clearly, the presence of the above described sub-clustering coupled with the
large number of LMC field stars located near NGC 1850 is going to complicate
attempts to obtain reliable surface photometry. Thus, it is advantageous to
remove the luminosity contribution of any definite cluster non-members.
DoPHOT was used to remove the stars in regions 1 and 3 (see Fig. 8) from
both the B and V frames.

It is also favorable to determine the cluster ellipticity so that
correspondingly elliptical apertures can then be used to perform the surface
photometry. This is a difficult task for most globular clusters and
particularly so for NGC 1850.  In common with other clusters, NGC 1850
consists of resolved stars, but in this case the stars dominating the light
are young supergiants. As well, there is the presence of the young binary
companion.  Fig. 10 shows a contour plot of the B and V star-subtracted,
median filtered (filter radius of 9\sec) images (solid lines). Superimposed
on the contours are ellipses produced by the ELLIPSE task in the
IRAF\footnote{$^1$}{IRAF is distributed by the National Optical Astronomy
Observatories, which is operated by the Association of Universities for
Research in Astronomy, Inc., under contract to the National Science
Foundation.} STSDAS package which uses the ellipse fitting technique of
Jedrzejewski (1987). Three things are immediately clear: 1) The ellipses do
not provide a good model for the smoothed NGC 1850 light distribution, 2)
the elliptical parameters change rapidly as a function of radius, and 3)
there is rather poor agreement between the two different bandpasses. This
leads us to believe that the ellipticities which we are measuring probably
result from the presence of a number of bright stars which could not
be adequately subtracted due to the extreme crowding. They are, therefore,
not a good representation of the shape of the underlying mass distribution
and hence, lacking anything better, we use circular contours to perform the
surface photometry. We cannot, however, rule out the possibility that NGC
1850 is significantly elliptical. The use of circular apertures in such a
case does not tend to result in systematic errors in the surface photometry,
but does increase the photometric scatter (i.e., see Fischer \et 1992b).

The cluster center was found using a multi-step procedure. The first step
was to estimate the position of the center using the isophotes produced
above.  Surface photometry was then obtained utilizing this cluster center.
A King model (see \S 3) was fit to the surface photometry and a new image
was made with the fitted King parameters but with a smooth light
distribution. We then cross correlated the artificial image with the
original image to obtain small residual $X$ and $Y$ shifts. The final
centers were within 2 arcsec of the original guesses.

Surface photometry was performed in a manner similar to Djorgovski (1988).
The frames were broken up into a series of concentric circular annuli
centered on the cluster.  The annuli were further divided into eight
azimuthal sectors. The average pixel brightness was determined for each
sector in a given annulus and the {\it median} of the eight separate
measurements was taken as the representative brightness at the area-weighted
average projected radius of the annulus (i.e., the mean radius of all the
pixels within the annulus which is approximately equal to the geometric
mean). The standard error of the median of the eight sectors was adopted as
the photometric uncertainty. Using the median as opposed to the more
commonly adopted mean is essential in the case of NGC 1850 as it reduces
the contamination due to the bright supergiants and the binary companion.

A background level (a combination of sky light and Galactic foreground and
remaining LMC field stars) was estimated from regions at large projected
distances from the cluster.  We found that the surface brightness profiles
tended to level out beyond 45 pc (we have adopted a distance to the cluster
of 50 kpc) for both the B and V frames. By ``levelling out'' we don't
necessarily mean that the cluster light does not extend beyond this point
but simply that fluctuations in the background dominate to such an extent
that it is no longer possible to observe the profile declining in intensity.
Therefore, it was this region with a projected radius of 45 $\le$ R (pc)
$\le$ 100 that was used for the background determinations. The
reddening-corrected background values for the B and V frames were,
respectively, $1370 \pm 27$ L$_B$\sol\ pc$^{-2}$ and $682
\pm 14$ L$_V$\sol\ pc$^{-2}$.

The background-subtracted surface photometry data is presented in Table 2
[assuming a cluster distance of 50 kpc, M$_{V\odot} = 4.83$ and (B-V)$_\odot
= 0.65$, Mihalas and Binney 1981, p. 60]. Columns 1 and 3 are the projected
area-weighted radii, and columns 2 and 4 are the B and V luminosity
densities, respectively.

Fig. 11 is a plot of the B and V surface brightness profiles. Also shown are
typical stellar profiles which have a FWHM less than 15\% of the cluster
core radius (see
\S 3); hence seeing will have a negligible effect on measurements of this
quantity (Mihalas and Binney 1981, p. 315).

\medskip
\centerline{\it 3. KING-MICHIE MODELS FOR THE SURFACE PHOTOMETRY}
\medskip

We fit projected single-component King-Michie (KM) models (King 1966 and
Michie 1963) to the surface photometry data. These models have an energy
($E$) angular momentum ($J$) per unit mass distribution function given by
$$f(E=-0.5v^2+W,J)
\propto e^{-[J/(2v_sr_a)]^2}(e^{-0.5v^2+W}-1),\eqno(\equat)$$ where $v_s$ is
the scale velocity, $r_a$ is the anisotropy radius (both described below),
and $W$ is the reduced gravitational potential. The shape of the density
distribution, $\rho_K$(r), is determined by solving Poisson's equation and
is dependent on two parameters; the central potential W$_\circ$, and $r_a$,
beyond which stellar orbits become increasingly radial. Scaling is applied
in both the radial ($r_s$) and luminosity ($\rho_{K\circ}$) dimensions to
give the best fit. For a complete description of the models (albeit for the
more complex multi-component case) the reader is directed to Gunn and
Griffin (1979).  Model density profiles with $r_a$ values ranging from 3
$r_s$ to infinity (i.e., an isotropic distribution function) were generated,
and projected on to the observational plane. These were binned identically
to the observed data and fit using a maximum likelihood technique.

Tables 3 and 4 show, for each $r_a$, the best $fitted$ KM parameters for both
bandpasses.  Column 1 contains $r_a$, column 2 is the reduced central
potential, column 3 is $r_s$ and column 4 is c = $r_s$/$r_t$ ($r_t$ is the
tidal radius).  Column 5 is the reduced chi-squared ($\chi_\nu^2 =
\chi^2/\nu$, where $\nu = 18$ is the number of degrees of freedom) for the
fit. Column 6 is the probability of obtaining a value greater than
$\chi^2_\nu$ for a model with the given parameters and the uncertainties
listed in Table 2.  These were derived from 1000 simulations of the surface
photometry data per model.  Each simulation used a surface profile generated
from the best fit model with errors, drawn from the uncertainties shown in
Table 2.  The same fitting procedure originally applied to the real data was
utilized and, in this way, we found the uncertainties in each fitted
parameter as well as the distribution of $\chi_\nu^2$ . The remaining
columns of Tables 3 and 4 will be discussed in \S 5.1. The fitted KM model
parameters were consistent for the two bandpasses.

Tables 5 and 6 have a summary of the $derived$ KM parameters for both
bandpasses corresponding to the models specified in Tables 3 and 4. Column 1
is the anisotropy radius while column 2 is the central luminosity density,
$\rho_{K\circ}$. Column 3 is the cluster luminosity and columns 4 and 5 will
be discussed in \S 5.1.

The solid lines in Fig. 11 show the best fit models which are those
possessing isotropic orbits. The quality of the fits deteriorates as $r_a$
decreases.  Interestingly, this is quite different from what was seen in the
case of the slightly older LMC cluster NGC 1866 which favored highly
anisotropic (i.e.  $r_a$ = 3 $r_s$) KM models and appeared to have a halo
of unbound stars (Fischer \et 1992a).  Also shown as dotted lines in the
same figure are the $r_a$ = 3 $r_s$ models. It is unclear whether or not the
data exhibit a tidal cut-off. In fact, using a profile that extends out to
about 20 pc, Elson (1991) finds no evidence for truncation. Our profile
extends to about twice that radius and there is, at best, a small indication
for truncation.  To test the hypothesis that there is significant truncation
we have fit models of the form: $$ \mu_ (R) = \mu_\circ [1 +
(R/a)^2]^{-\gamma/2},\eqno(\equat)$$ to the surface brightness profile using
a non-linear, weighted, least squares technique.  The best-fit parameters
for these untruncated power-law models are:  $\mu_{B\circ} = 26800.
\pm 2000.$\sol\ pc$^{-2}$, $a_B = 2.8 \pm 0.3$ pc, $\gamma_B = 2.26 \pm 15$,
$\mu_{V\circ} = 12000. \pm 900.$\sol\ pc$^{-2}$, $a_V = 3.1 \pm 0.4$ pc,
$\gamma_B = 2.34 \pm 0.15$. The $\chi^2_\nu$ are 0.89 [P($> \chi^2_\nu$)
0.59] and 1.06 [P($> \chi^2_\nu$) 0.39] for the B and V profiles,
respectively, which are higher than all but the $r_a$ = 3 $r_s$ KM models.
We conclude that the best truncated models provide marginally better
agreement with the surface photometry data.

\centerline{4. RADIAL VELOCITIES}
\medskip

\centerline{\it 4.1) Observations and Reductions}
\medskip

Spectra of 52 supergiants in the region surrounding NGC 1850 were obtained
during two runs (1991 February 14-20 and 1991 December 14-17) using the
photon-counting echelle spectrograph on the 2.5m Dupont reflector, designed
and built by Steve Shectman. Eight of the stars have repeat measurements.

The observation and reduction procedures for a previous run at LCO have been
discussed extensively in Welch \et (1991) and remain largely unchanged for
this data. Briefly, the observing procedure consisted of exposures with
integration times of 200 - 500s and Th-Ar arcs approximately every 45
minutes. A representative LCO spectrum is shown in Fig. 2 of C\^ot\'e \et
(1991).  The reduction utilizes the IRAF ECHELLE and RV packages (Tody 1986)
to obtain both velocities and velocity uncertainties. The velocity
zero-point is tied to the IAU velocity standard 33 Sex as described in
Fischer \et 1992a and is believed to be accurate to better than 2 \kms.

Along with the program stars, relatively high S/N spectra were obtained on
each night for both a nearby bright star used as a local velocity standard
(LVS) for differential measurements and the radial velocity standard HD
23214 to examine the possibility of velocity zero-point drifts and to test
the accuracy of the velocity uncertainties returned by RVXCOR.  These
velocities are shown in Table 7 for the eight separate nights over the two
observing runs in which NGC 1850 stars were observed.  Column 1 is the
Heliocentric Julian Date minus 2448000, column 2 is the radial velocity,
column 3 is the mean velocity for the given observing run and column 4 is
the reduced chi squared for the appropriate observing run. Within the RVXCOR
produced uncertainties, there is no significant zero-point drift during the
individual runs. There is, however, a discrepancy between the mean
velocities for the LVS for the two separate observing runs at the
4.6$\sigma$ level implying either that the radial velocity zero-point has
shifted or that this star has a variable velocity.  The shift is not seen in
HD 23214 implying the latter hypothesis is probably correct.  Olszewski \et
(1991) have obtained 9 lower resolution radial velocity measurements for HD
23214 during two observing runs, 1987 January 8-12 and 1987 December 27-31,
deriving a mean velocity of -4.3 $\pm$ 1.8
\kms. They also present a single CORAVEL velocity determination of -4.7 $\pm$
0.3 \kms\ for HJD = 2446862.51.  This latter deviates by 5.6$\sigma$ from
our value of -2.8 $\pm$ 0.2 \kms, which, as mentioned above, is based upon
the IAU radial velocity standard 33 Sex.  Another radial velocity standard,
HD 196983, has been measured at two epochs, HJD -- 2448000 = 606.5225 and
608.5179 yielding v$_x$ = --9.6 $\pm$ 0.5 \kms\ and --8.7 $\pm$ 0.6 \kms,
respectively, for a mean of --9.2 $\pm$ 0.4 \kms.  This agrees very well with
the CORAVEL value, based on 10 separate measurements spanning 715 days
starting in September 1981, of --9.3 $\pm$ 0.1
\kms\ (Maurice \et 1984). To conclude, we feel that our velocity zero-point
is uncertain at about the 2.0 \kms\ but that there is no significant
zero-point drift during the two runs. (This latter possibility will be
examined in more detail presently.)

The radial velocity data for the cluster stars are presented in Table 8.
Column 1 contains the stellar identifications (these do not correspond to
the ID's in Table 1), column 2 has the projected radii, column 3 the equinox
J2000.0 position angles, column 4 contains the radial velocities and column
5 contains the mean velocities for stars with repeated measurements.  Column
6 is the Heliocentric Julian Date -- 2448000 for the velocity measurements.
Columns 7 and 8 are V and B--V for the stars.  The photometry in the
innermost regions is relatively uncertain due to the high degree of
crowding. Fig. 12 is a finder chart for all but the star farthest from the
cluster center.

Of the eight stars with repeated velocity measurements, six had individual
$\chi^2$ of less than 2.85 for a total $\chi^2 = 6.44$ for 5 degrees of
freedom. Of the other two, one had $\chi^2 = 7.9$ (star RV 49) and the other
had $\chi^2 = 217$ (star RV 19). If the velocity shift indicated by the LVS
is applied these $\chi^2$ values all decrease; we get a total $\chi^2 = 4.33$
for the six stars and 3.40 and 217.00 for the other two stars. This is an
argument in favor of adoption of the velocity shift.

{}From their colors, the two high-$\chi^2$ stars are evolved and their
velocities indicate that they are in the LMC. Star RV49 appears to be an LMC
field star, perhaps a binary. Star RV19, as indicated by its radial
velocities, is clearly a member of the LMC. Further, its luminosity (V =
13.54) indicates that it is a supergiant. Two possibilities are that the
star is a binary or a Cepheid. The two velocity measurements differ by over
30 \kms\ which argues against the former hypothesis as it would require a
very favorable inclination coupled with fortuitous observations. The stars
color (B -- V = 0.65) makes it an excellent Cepheid candidate, and it is
likely to be a cluster member based on its proximity to the cluster center
(\Rc2 = 3.8 pc). Neither of these two stars will be used in the cluster mass
determinations.

In Fig. 13 we present plots of radial velocity and radial velocity
uncertainty vs. stellar B--V. There are clearly trends towards both a larger
apparent velocity dispersion and larger uncertainties for the bluer stars.
The correlation between velocity uncertainty and apparent velocity
dispersion is, of course, expected. The correlation between uncertainty and
color has two causes: 1) the lack of lines in the bluer stars, and 2)
spectral mismatch with the template star (a K giant). For the mass
determinations of the next section we decided not to employ any stars with
uncertainties greater than 2.7 \kms\ (the solid lines in Fig. 13), eliminating
9 stars. We have also eliminated the three stars having v$_x > 270$ \kms\ as
being probable non-members (or variables) and star RV 18, which was classified
as a variable by Robertson (1974) and from its color is a probable cepheid,
leaving a total of 38 stars.

\medskip
\centerline{5. VELOCITY DISTRIBUTIONS AND MASS DETERMINATIONS}
\medskip

\medskip
\centerline{\it 5.1 King-Michie Models}
\medskip

The mass of a multi-mass KM model is given by $$M = {9r_sv_s^2 \over 4\pi G}
\int{{\rho \over \rho_0}r^2dr}\eqno(\equat)$$ Illingworth (1976), where
$r_s$ is given in Tables 3 and 4, and $v_s$ is the scale velocity. The run
of $\sigma^2_r(r)$ and $\sigma^2_t(r)$ are determined from
$$\sigma_{r,t}^2(r) = {\int_{|\sigma|
\le W(r)} f(\sigma,W)\sigma_k^2d^3\vec \sigma \over \int_{|\sigma|
\le W(r)} f(\sigma,W)d^3\vec\sigma},\eqno(\equat)$$ where $W$ is the reduced
potential (W = 0 at the tidal radius) and $\sigma_k = \sigma$cos$\theta$ or
$\sigma$sin$\theta$ for $\sigma_r$ or $\sigma_t$, respectively. Comparisons
were made between the observed velocities and the model velocity dispersion
projected along the line of sight, $$\sigma_p^2(R) = {2 \over \mu (R)}
\int^\infty_R{\rho_K(r)[(r^2-R^2)\sigma_r^2(r)+R^2\sigma_t^2(r)]dr \over
r(r^2-R^2)^{1/2}},\eqno(\equat)$$ (Binney and Tremaine 1987, p. 208),
yielding the optimal value for $v_s$. The comparison was accomplished using
the maximum likelihood technique outlined in Gunn and Griffin (1979). Simply
put, the probability density function for v$_{x i}$, an observed stellar
velocity, is a Gaussian with standard deviation equal to the model
dispersion plus the velocity uncertainty added in quadrature:  $$P_i \sim {1
\over
\sqrt{v_{err\ i}^2+v_s^2v_{p\ i}}} e^{-(v_{x\ i} - v_{ave})^2/2(v_s^2 v_{p\
i}^2 + v_{err\ i}^2)}.\eqno(\equat)$$ One minimizes this function with
respect to $v_s$ and $v_{ave}$ resulting in two equations which can be
solved simultaneously for the most probable values of the two parameters.

A serious problem in mass determinations is contamination from binary and
non-member stars, both of which tend to increase the mass estimate. This is
particularly a problem when one has a relatively small sample of stars and a
small velocity dispersion as is the case for NGC 1850. A single interloper
in the data sample can easily result in a 50\% or greater overestimate in
the mass.  We attempted to deal with this problem in the following manner.
First, using the entire data set, the optimal $v_s$ and $v_{ave}$ are
determined using equation 6. For every star the parameter $$\delta_i =
\sqrt{{(v_{ri}-v_{ave})^2 \over {v_s^2
\sigma_p^2(R_i) + v_{err\ i}^2}}} \eqno(\equat)$$ is tabulated. The star
with the largest $\delta_i$ is removed and the procedure is repeated until
all the stars have been removed. Then, using the KM models, we performed
Monte Carlo simulations of the radial velocity data. We started with the
known projected radii ($R_i$) of the program stars. The true radius is in the
range \Rc2 $\le$ r $\le$ r$_{max}$, where r$_{max}$ can extend to infinity for
an unbound distribution. If $x$ is the displacement from the mean cluster
position along the line-of-sight such that r = $\sqrt{R^2+x^2}$ then the
probability that the star is at $x$ is $$p(x) \sim
\rho_K(\sqrt{R^2+x^2}).\eqno(\equat)$$ A three-dimensional position along
with corresponding model-dependent radial and tangential velocities were
drawn at random from their respective probability distributions. The
velocity component along the line-of-sight was then determined, and an error
term, drawn from a Gaussian distribution with standard deviation equal to
the velocity error, as tabulated in Table 8, was added.  This process was
repeated, producing 10000 sets of data, each with a given mass and $r_a$ and
the same projected positions and velocity measurement errors as the original
data set. Finally the maximum likelihood technique was applied to each of
the artificial data sets and the maximum $\delta_i$ were recorded. The $v_s$
for the first three iterations (using the unshifted radial velocity data) are
shown in column 2 of Table 9. Columns 3 and 4 are the maximum $\delta_i$ and
the percentage of simulations with $\delta_{max}$ exceeding this value,
respectively, and column 5 is the star possessing $\delta_{max}$. A value of
zero in the fourth column indicates less than 0.001. We feel that there is a
fairly high probability that the first two stars are either variables or
non-members and hence they will not be used for the mass determinations. It
is worth remembering, however, that their removal results in a mass
reduction of about 80\% demonstrating the extreme sensitivity of the mass
determinations to interlopers. We repeated the procedure with data that was
corrected according to the LVS (see \S 4.1) and found similar results but
obtained a $v_s$ that was about 2\% larger. The fact that it causes an
increase in the velocity dispersion argues against its use, but in any case
it does not have significant repercussions on the mass estimates and hence
was not adopted.

Fig. 14 shows mean radial velocity vs. projected radius (upper panel) and
versus position angle (lower panel) for the 36 remaining stars. The solid
lines are the mean velocity, $\vave = 251.4 \pm 2.0$ km s$^{-1}$.

The values of $v_s$ obtained from this reduced data set are displayed in
column 7 of Tables 3 and 4. The corresponding masses and M/L ratios are in
columns 4 and 5 of Tables 5 and 6. One can see that the total
cluster mass and M/L are fairly insensitive to assumptions about $r_a$, and
the best values are M = 5.7 $\pm 2.3 \times 10^4$ M\sol\ and M/L$_B$ = 0.02
$\pm 0.01$ M\sol/L$_B$\sol\ or M/L$_V$ = 0.05 $\pm 0.02$ M\sol/L$_V$\sol.

The above-described Monte-Carlo orbit simulations were used to determine the
uncertainties implicit in the maximum likelihood technique and to search for
any possible systematic effects.  We found that this method tended to
underestimate $v_s$ by about 3\% and hence the mass by 6\%. The values in
the relevant tables have been corrected for this effect and the
uncertainties shown were derived from the simulations.

A goodness-of-fit statistic $$\zeta^2 = \sum{{(v_{x\ i} - v_{ave})^2} \over
(v_s^2 v_{p\ i}^2 + v_{err\ i}^2)}\eqno(\equat)$$ was generated for each
value of $r_a$ and is shown in column 8 of Tables 3 and 4 (34 degrees of
freedom).  The distribution of this statistic can be extracted from the
Monte Carlo simulations. Column 9 shows the probability of exceeding the
observed $\zeta^2$ assuming that the cluster velocities are specified by the
model parameters indicated and have the uncertainties tabulated in Table 8.
The isotropic models yielded the best agreement with the data (although, only
marginally so) consistent with the findings from the surface photometry.

Because of the similarity between the velocity dispersion and the velocity
uncertainty, the mass estimates are quite highly dependent on the the
accuracy of the uncertainty estimates. We have already discussed the
analysis of the stars with repeated measurements and concluded that the
uncertainties seem to be reasonable with reduced $\chi^2$ near unity.
However, it is worth quantifying the effects of the uncertainties.  Assuming
that the uncertainties are actually zero causes and increase in the estimates
of the $v_s$'s by approximately 25\%, or a roughly 50\% increase in cluster
mass. Increasing the uncertainties by 50\% lowers the $v_s$'s by 50\% and
the mass by 75\%.

\medskip
\centerline{\it 5.2 Rotation}
\medskip

A careful examination of the radial velocities vs. position angle (Fig. 14)
reveals evidence for a sinusoidal variation which may be indicative of
cluster rotation. In order to test this hypothesis we measured the
difference in median velocities on either side of an imaginary axis which is
stepped around the cluster center at 1$^\circ$ intervals. If one is viewing
an edge-on rotating system then a sinusoidal variation is expected with the
velocity difference maximized when the axis corresponds to the rotation
axis. This effect will degrade as the system deviates from edge-on.  The
bottom left panel of Fig 15 shows a plot of the velocity differences versus
the position angle of the axis. The best-fit sine curve (using unweighted
least-squares) has an amplitude of A = 2.1 km s$^{-1}$ and the implied
projected rotation axis is 100$^\circ \pm 40^\circ$ ($\chi^2 = 25.12$ for 178
degrees of freedom). To test the significance, we constructed 1000
non-rotating models as described above and performed the same test.  Only
7\% of these models had amplitudes exceeding A = 2.1 and hence we feel
fairly confident that NGC 1850 is rotating. Rotation has previously been
detected at the 97\% confidence level for a sample of 69 stars in the young
LMC cluster NGC 1866 (Fischer \et 1992a) which had A = 1.8 \kms\ and $v_s =
3.1$ \kms. In that case, incorporation of rotation resulted in only a very
small downward changes in the mass estimate. However, in the case of NGC
1850, the two parameters have very similar values meaning rotation may be
dynamically more significant than it was in NGC 1866.

Unfortunately, as explained in \S 2.3 it is impossible to extract a meaningful
ellipticity from the NGC 1850 light distribution. This information would
have enabled us to construct rotating oblate cluster models to compare with
the radial velocities. We therefore adopted an alternative approach of
assuming different ellipticities values for the cluster, constructing models
for the rotation, and comparing them to the data.

For an axisymmetric system, the relevant Jeans' equations (velocity moments
of the collisionless Boltzmann equation) in cylindrical coordinates are:
$${\partial(\rho \sigma_{\hbox{\fsmall R}}^2) \over \partial \hbox{R}} +
{\partial(\rho\sigma_{\hbox{\fsmall R}z}) \over \partial z} +
\rho\left({\sigma_{\hbox{\fsmall R}}^2 - \sigma_\phi^2 - v_\phi^2 \over
\hbox{R}}\right) + \rho{\partial\Phi \over \partial \hbox{R}} =
0,\eqno(\equat)$$ and $${\partial(\rho\sigma_{\hbox{\fsmall R}z}) \over
\partial \hbox{R}} + {\partial(\rho\sigma_z^2) \over \partial z} +
{\rho\sigma_{{\hbox{\fsmall R}}z}
\over \hbox{R}} + \rho{\partial\Phi \over \partial z} = 0,\eqno(\equat)$$
where (R,$\phi$,z) are the cylindrical coordinate axes, the
($\sigma_{\hbox{\fsmall R}}, \sigma_\phi,
\sigma_z$) are the corresponding velocity dispersions, and $v_\phi$ is the
rotation velocity. $\Phi$ is the gravitational potential.

Both the rotating and non-rotating models which were used have velocity
ellipsoids aligned with the cylindrical coordinate axes (i.e.
$\sigma_{\hbox{\fsmall R}z}=0$) and, as well, both have $$\sigma_\phi =
{\sigma_{\hbox{\fsmall R}}
\over \sqrt{1 + (\hbox{R}/\hbox{R}_a)^2}},\eqno(\equat)$$ where $\hbox{R}_a$
can be varied up to $\infty$. The rotating models also have the
condition $$\sigma_{\hbox{\fsmall R}} = \sigma_z,\eqno(\equat)$$ implying
$$v_\phi^2 =
\hbox{R}{\partial\Phi \over
\partial \hbox{R}} + {\hbox{R} \over \rho}{\partial \over \partial
\hbox{R}}\int_z^\infty{\rho
{\partial\Phi \over \partial z}dz} + {1 \over
\rho}\left[1 - {1 \over
\sqrt{1 + (\hbox{R}/\hbox{R}_a)^2}}\right]\int_z^\infty{\rho{\partial\Phi \over
\partial
z}dz}.\eqno(\equat)$$

The models are constructed by assuming that the mass distribution is
equivalent to the deprojected light distribution (constant M/L).  Equations
11 and 14 can then be solved directly to obtain $\sigma_z$, and $v_\phi$.
Once $v_\phi$ is known it can be substituted into equation 10 which, in turn,
can be solved for $\sigma_{\hbox{\fsmall R}}$ and $\sigma_\phi$. This is
outlined in Binney and Tremaine (1987, p. 209) for the $\sigma_{\hbox{\fsmall
R}} = \sigma_\phi = \sigma_z$ case and a $\epsilon = 0.3$ model is shown in
Fig. 6 of Fischer \et (1992b).

Once the models have been generated, it is simply a matter of projecting and
then scaling them using a similar maximum likelihood method to that employed
for the KM models. For our purposes we have constructed models spanning
$\epsilon = 0.1 - 0.3$ for R$_a = \infty$. All of the models assume a cluster
inclination of 90\deg.  Table 10 displays the results of the model fitting.
Column 1 is the ellipticity, column 2 the cluster mass, columns 3 and 4 the
B and V M/L's, and columns 5 and 6 are $\zeta^2$ and P($>\zeta^2$) as above.
These models all have P($>\zeta^2$) greater than or equal to the best
non-rotating KM models despite having larger $\zeta^2$ values. However, as
can be seen from Fig. 7 of Fischer \et (1992b) the rotating models have a
wider $\zeta^2$ distribution. The rotating models have masses (and M/L's)
which are marginally lower than the non-rotating models. Fig. 15 exhibits
plots of the velocity differences vs. axis position angle for the original
and rotation-subtracted data. Column 7 of Table 10 is the amplitude of the
best-fit sine curves for each case. One can see that the amplitude decreases
for the rotation subtracted data with increasing $\epsilon$. It does not,
however, decrease smoothly to zero and one can see that for the $\epsilon =
0.3$ case there is a residual trend in the data which does not disappear for
higher ellipticities. This probably results from some type of model mismatch
which may be a result of the model itself or might be due to a poorly
determined rotation axis position angle or an inclination of less than
90\deg. We changed the rotation axis position angle by plus and minus
20$^\circ$ but found that the residual velocities did not improve. Another
possibility is that the characteristic rotational signature arises from the
interaction between the two binary cluster components. The resulting stellar
motions might in such a case have a similar appearance to rotation.

\medskip
\centerline{6. CONSTRAINTS ON THE MASS FUNCTION}
\medskip

The cluster M/L estimate can be used to constrain the slope of the initial
mass function (IMF). In this study we used an IMF of the form $$\phi(m) =
m^{-(x+1)} ~dm~~~~~~~m \ge m_d,\eqno(\equat)$$ $$\phi(m) = m
{}~dm~~~~~~~~~~~~~ m < m_d,\eqno(\equat),$$ which gives a drop-off at the
faint end similar to what is seen in the solar neighborhood (Miller
\& Scalo 1979).

The theoretical cluster M/L is given by $${M \over L} =
{\int^{m_u}_{m_{l}}m\phi(m)dm \over
\int^{m_u}_{m_{l}}l(m)\phi(m)dm},\eqno(\equat)$$ where $l(m)$ is the
luminosity of a star of mass $m$ given by a theoretical mass-luminosity
relationship for main-sequence and evolved stars. We used the
mass-luminosity relationship of Mermilliod (1992), supplemented with the
Bergbusch \& Vandenberg (1992) values in the range 0.15 $\le m$ (M$_\odot)
\le$ 0.9, with the cluster parameters determined in \S 2.2. The initial mass
of a star now at the end of the asymptotic giant branch is approximately 5.4
M\sol. We have adopted the treatment of Pryor \et 1986 to deal with
remnants: stars with initial masses of 5.4 - 8.0 M\sol\ become white dwarfs
with masses of 1.2 M\sol\ and stars with initial masses greater than 8
M\sol\ are assumed to be ejected from the cluster as is consistent with the
high velocities seen for pulsars in the disk (Gunn and Griffin 1979).
Therefore, the choice of m$_u$ is not important except for estimating the
amount of mass lost from the cluster due to stellar evolution.

Table 11 shows the derived values of $x$ (column 3 and 5) for 3 different
$m_l$ (column 1), and 5 different $m_d$ (column 2).  Columns 4 and 6 contain
the implied mean stellar mass for the given model. The values of $x$
correspond to the B and V M/L's indicated at the top of the table. Although
the values corresponding to the B and V M/L's are consistent within the
uncertainty, the B-band $x$'s are systematically lower. It appears that for
a given set of assumptions the V-band M/L constrains the mass function slope
somewhat more tightly than the B-band. Furthermore, the V-band is also less
sensitive to the assumed cluster age although one finds that the both the
slopes steepen for a younger assumed age and flatten for older (this has not
been accounted for in the uncertainty). The B-band has one advantage in that
it appears to be less sensitive to the low mass cut-off.Both sets of slopes
are substantially shallower than was seen for the young LMC cluster NGC 1866
(Fischer \et 1992a), which had an average slope (i.e. as determined without
employing a drop-off in the mass function) of between $x$ = 1.35 $\pm 0.1$
and $x$ = 1.82 $\pm 0.1$ (depending on adopted age) for $m_l = 0.1$.

\medskip
\centerline{7. Evolutionary Timescales, Mass Segregation and Equipartition of
Energy}
\medskip

The stars with measured velocities are all supergiants, with relatively
large masses. If equipartition of energy has occurred these stars, being the
most massive, would have a velocity dispersion below the mean value for all
mass classes. Therefore, in order to justify the use of single-mass models
we must demonstrate that the cluster is sufficiently young such that
equipartition and mass segregation have not had enough time to become
significant effects.  Two important timescales are the central relaxation
time $$t_{r\circ} = (1.55 \times 10^7
\hbox{yr})\left({r_s
\over \hbox{pc}}\right)^2\left({v_s \over \hbox{km s}^{-1}}\right)
\left({M_\odot
\over \left< m \right>} \right)\left[\log(0.5 M / \left< m
\right>)\right]^{-1} = 0.4 - 2.2 \times 10^8 \hbox{yr}\eqno(\equat)$$
(Lightman and Shapiro 1978) and the half mass relaxation time $$t_{rh}=(8.92
\times 10^8 \hbox{yr})\left({M
\over 10^6 M_\odot}\right)^{1/2}\left({r_h \over \hbox{pc}
}\right)^{3/2}\left({M_\odot \over
\left< m \right>}\right)\left[\log(0.4M/\left< m \right>)\right]^{-1} = 1.5
- 5.3 \times 10^9 \hbox{yr}\eqno(\equat)$$ (Spitzer and Hart 1971). The
parameters in these equations have all been discussed previously except for
the half mass radius, $r_h \simeq 11$ pc. Aside from the inner core the
cluster is in a dynamically unevolved state and, therefore, we do not expect
substantial energy transfer to have occurred. We conclude that no large
systematic errors are being introduced into the velocity dispersions.

Another possible problem is primordial mass segregation. This would occur if
star formation in the dense core region favored a different ratio of
high-to-low mass stars than at larger radii. While this should not effect the
velocity dispersion greatly, it would mean that the assumption of a uniform
M/L is incorrect. Consequently the derived luminosity density profiles would
not be an adequate representation of the mass density profile. One should be
able to construct luminosity functions at different projected cluster radii
and hence search for a gradient in mass function. In practice, owing to the
very crowded nature of the inner cluster regions and the high surface
density of non-member stars present on the frame this would be difficult to
accomplish in a convincing manner and we have not attempted it.

\medskip
\centerline{VII. CONCLUSIONS}
\medskip

In this paper we have examined the age and internal dynamics of the young
binary LMC cluster NGC 1850 using BV CCD images and echelle spectra of 52
supergiants.

\itemn A BV CMD was constructed for the field surrounding the cluster and
was found to contain 3 distinct populations of stars. The first was a very
young population of age $\tau = 6 \pm 5$ Myr belonging primarily to the
smaller member of the binary system located 30\sec\ west of the larger
member. This young population allowed for an accurate reddening estimate of
E(B-V) = 0.17 $\pm 0.03$ mag. The second, slightly older population, belongs
primarily to the the larger cluster and has an age of $\tau = 90 \pm 30$
Myr. The third population was older and mainly comprised of LMC field stars.

\itemn Attempts were made to determine ellipticity parameters for the
cluster using star-subtracted, median filtered BV images. This was greatly
complicated by the presence of extremely bright young resolved stars and the
binary nature of the cluster and no meaningful shape parameters were derived.

\itemn BV luminosity profiles were constructed out to projected radii of R
$>$ 40 pc.  Single component King-Michie (KM) models were applied to this
data in order to determine the most favorable value for the anisotropy
radius and the total cluster luminosity. The luminosity varied from L$_B$ =
2.60 - 2.65 $\pm 0.2 \times 10^6$ L$_B$\sol\ and L$_V$ = 1.25 - 1.35 $\pm
0.1 \times 10^6$ L$_V$\sol\ as $r_a$ went from to infinity to 3$r_s$. The
fitted and derived KM parameters for both bandpasses were consistent and the
isotropic models provided the best agreement with the data.

\itemn To test for the presence of a tidal cut-off in the luminosity
profile, a power-law model without truncation was applied to the data. This
model was found to provide marginally worse fits than the KM models giving
some indication that the cluster density distribution is truncated.

\itemn Of the 52 stars with echelle spectra, a subset of 36 were used to
study the cluster dynamics. The KM radial velocity distributions were fitted
to these velocities yielding scale velocities of $v_s = 2.2 - 2.3 \pm 0.5$
\kms\ for the $r_a$ range employed. The total cluster mass was 5.4 - 5.9 $\pm
2.4 \times 10^4$ M\sol\ corresponding to M/L$_B$ = 0.02 $\pm 0.01$
M\sol/L$_B$\sol\ or M/L$_V$ = 0.05 $\pm 0.02$ M\sol/L$_V$\sol. The mean
cluster velocity is $\vave = 251.4 \pm 2.0$ km s$^{-1}$.

\itemn A rotational signal in the radial velocities has been detected at the
93\% confidence level implying a rotation axis at a position angle of
100\deg $\pm 40$\deg. A variety of rotating models were fit to the velocity
data assuming
$\epsilon = 0.1 - 0.3$.  These models provided slightly better agreement
with the radial velocity data than the KM models and had masses that were
systematically lower by a few percent.

\itemn Values for the slope of the mass function were determined using the
derived M/L, theoretical mass-luminosity relationships, and several forms
for the IMF. The preferred value for the slope of a power-law IMF is a
relatively shallow, $x = 0.29 \pmm{+0.3}{-0.8}$ assuming the B-band M/L or
$x = 0.71 \pmm{+0.2}{-0.4}$ for the V-band.

\itemn The current cluster age is similar to its central relaxation time but
about 2 orders of magnitude less than its half-mass relaxation time.
Therefore, aside from in the inner core the cluster is in a dynamically
unevolved state and we expect that equipartition has not yet occurred in any
substantial way.

\bigskip
\centerline{ACKNOWLEDGEMENTS}
\medskip

P.F. would like to knowledge the Natural Sciences and Engineering Research
Council (NSERC) for a post-graduate fellowship. This work was undertaken
while D.L.W. was a NSERC University Research Fellow. Partial support for this
work was provided by NASA through grant \# HF-1007.01-90A awarded by the
Space Telescope Science Institute which is operated by the Association of
Universities for Research in Astronomy, Inc., for NASA under contract
NAS5-26555. We would like to thank Dr. Maeder, Dr. Meynet, Dr Schaller, and
Dr. Schaerer for a copy of their new isochrones and Dr. P. Bergbusch for a
copy of his isochrones, both, prior to publication.

Reprints of this paper are available through mail or anonymous FTP. Contact
Fischer@crocus.physics.mcmaster.ca for information.

\vfill\eject
\centerline{FIGURE CAPTIONS}

\figc A B band CCD image of NGC 1850 displayed to highlight fainter features.
The image is approximately 10\min square with north at the top and east to
the left. See text for an explanation of the numbers and circles.

\figc The same as Fig. 1 but displayed to show only the bright stars.

\figc Comparison between photometry from this work and Robertson (1974).
The zero-point differences can be found in the text.

\figc The same as figure 3 but for the photometry from Alcaino \& Liller
(1987).

\figc The same as figure 3 but for the photometry from (Elson 1991)

\figc BV color-magnitude diagram for the entire 10\min field surrounding
NGC 1850. The numbers correspond to three distinct populations (see text).

\figc BV color-magnitude diagram for the region within 12\sec of center of
the companion cluster (see Fig. 2). The solid line corresponds to the
isochrone with z = 0.008 while the broken line has z = 0.001. Both sets of
isochrones have ages of $\tau = 6$ Myr, as well as (m--M)$_\circ$ = 18.5 and
E(B--V) = 0.17 mag.

\figc BV color-magnitude diagram for the region within 1\min of the center
of the main cluster. The solid lines correspond to the isochrones with
z = 0.008 while the broken lines have z = 0.001. Both sets of isochrones have
ages of $\tau = 6$ Myr and $\tau = 90$ Myr, as well as (m--M)$_\circ$ = 18.5
and E(B--V) = 0.17 mag.

\figc Population synthesis Monte Carlo experiments for the z = 0.008 and z =
0.001 isochrones described in the previous figure. A photometric uncertainty
of 0.05 mag has been added.

\figc Contour plot of star-subtracted, median-filtered BV images of NGC 1850.
The solid lines are isophotes while the dashed lines are the best-fit
ellipses.

\figc B and V surface brightness profiles. The long-dashed lines are typical
stellar profiles. The solid and short-dashed lines are isotropic and $r_a =
3 r_s$ single-mass King-Michie surface density models, respectively. The
solid horizontal lines indicate the background levels.

\figc A finder chart for 51 of the 52 stars for which we have radial
velocities.

\figc Radial velocity (upper panel) and velocity uncertainty (lower panel)
vs. B-V.

\figc Mean radial velocity vs. projected radius (upper panel) and versus
position angle (lower panel) for the 36 remaining stars. The solid lines are
the mean velocity, $\vave = 251.4 \pm 2.0$ km s$^{-1}$.

\figc The lower left panel displays the difference in median velocity for
stars on either side of an axis at the specified position angle. Also shown
is the best fit sine function corresponding to a rotation axis with position
angle 100$^\circ$. The other panels show the same thing but for velocity
data which has been rotation-subtracted assuming the specified ellipticities
and the models described in \S 5.2.
\vfill\supereject
\baselineskip=13pt
\pageinsert
\fsmall{
\tabhl{\taone}{BV Photometry}
\span\tcoli &
\span\tcol{.05} &
\span\tcol{.05} &
\span\tcol{.05} &
\span\tcoll{.05}{4.5}{3.5} &
\span\tcol{.1} &
\span\tcol{.05} &
\span\tcol{.05} &
\span\tcol{.05} &
\span\tcoll{.05}{4.5}{3.5} &&
\span\tcol{.1} \cr
\topperl

&&&&&&&&&\cr
\hfil ID & X & Y & V & B--V & ID & X & Y & V & B--V & ID & X & Y & V & B--V \cr
&&&&&&&&&\cr
\tablerule
&&&&&&&&&\cr
 !1 & 511.8 & 417.2 &18.04 &\phantom{-}1.04 &\phantom{-}51 & 505.7 & 444.8
&18.04 &-0.04 & 101 & 528.5 & 463.9 &13.40 &-0.14 \cr
 !2 & 528.8 & 418.0 &18.75 &-0.15 &\phantom{-}52 & 545.1 & 445.0 &18.54 &-0.14
& 102 & 498.6 & 464.4 &14.64 &\phantom{-}0.29 \cr
 !3 & 541.4 & 418.1 &16.37 &-0.00 &\phantom{-}53 & 485.8 & 445.1 &18.08
&\phantom{-}0.03 & 103 & 466.2 & 464.5 &16.84 &\phantom{-}1.58 \cr
 !4 & 502.5 & 419.7 &18.44 &\phantom{-}1.23 &\phantom{-}54 & 530.6 & 445.2
&14.79 &-0.14 & 104 & 548.6 & 464.5 &16.06 &\phantom{-}0.11 \cr
 !5 & 521.2 & 419.9 &16.28 &-0.19 &\phantom{-}55 & 550.8 & 445.7 &16.01
&\phantom{-}0.18 & 105 & 555.2 & 465.2 &18.49 &-0.03 \cr
 !6 & 545.2 & 420.4 &18.85 &-0.07 &\phantom{-}56 & 498.4 & 446.3 &18.30 &-0.10
& 106 & 481.9 & 465.3 &17.01 &-0.05 \cr
 !7 & 498.8 & 422.3 &17.07 &-0.02 &\phantom{-}57 & 492.8 & 446.5 &18.48
&\phantom{-}0.02 & 107 & 534.1 & 465.3 &14.01 &-0.13 \cr
 !8 & 506.3 & 423.8 &16.95 &-0.02 &\phantom{-}58 & 516.4 & 446.6 &16.42
&\phantom{-}0.05 & 108 & 518.2 & 465.9 &16.98 &-0.15 \cr
 !9 & 552.8 & 426.0 &15.90 &-0.05 &\phantom{-}59 & 463.3 & 447.1 &19.48 &-0.01
& 109 & 526.0 & 466.2 &15.40 &-0.16 \cr
 10 & 544.8 & 426.5 &16.82 &-0.08 &\phantom{-}60 & 579.3 & 447.2 &17.81 &-0.03
& 110 & 607.7 & 466.4 &17.95 &-0.11 \cr
&&&&&&&&&\cr
 11 & 558.7 & 426.6 &18.51 &-0.02 &\phantom{-}61 & 549.8 & 447.9 &16.87 &-0.11
& 111 & 485.1 & 466.5 &18.02 &-0.00 \cr
 12 & 533.7 & 426.8 &18.14 &\phantom{-}0.09 &\phantom{-}62 & 594.6 & 448.0
&18.78 &\phantom{-}0.00 & 112 & 542.2 & 466.8 &15.48 &-0.16 \cr
 13 & 480.4 & 427.3 &18.40 &-0.00 &\phantom{-}63 & 565.7 & 448.0 &18.81 &-0.19
& 113 & 610.3 & 467.1 &17.35 &\phantom{-}0.11 \cr
 14 & 522.6 & 429.1 &17.82 &-0.03 &\phantom{-}64 & 570.4 & 448.9 &18.53
&\phantom{-}0.38 & 114 & 480.2 & 467.4 &17.00 &\phantom{-}0.07 \cr
 15 & 499.4 & 429.4 &17.30 &-0.05 &\phantom{-}65 & 534.8 & 450.1 &18.18 &-0.08
& 115 & 477.9 & 467.6 &17.99 &-0.16 \cr
 16 & 572.9 & 429.4 &18.86 &\phantom{-}0.19 &\phantom{-}66 & 486.8 & 451.3
&17.99 &-0.07 & 116 & 562.6 & 468.7 &16.83 &-0.00 \cr
 17 & 531.8 & 429.4 &18.64 &-0.10 &\phantom{-}67 & 552.1 & 451.5 &19.08 &-0.14
& 117 & 459.1 & 468.8 &16.97 &-0.00 \cr
 18 & 565.5 & 430.2 &18.98 &-0.06 &\phantom{-}68 & 589.2 & 452.1 &17.85 &-0.06
& 118 & 511.2 & 469.1 &17.17 &-0.10 \cr
 19 & 502.7 & 430.2 &18.86 &-0.20 &\phantom{-}69 & 463.8 & 452.2 &17.17 &-0.01
& 119 & 534.6 & 469.5 &17.22 &-0.03 \cr
 20 & 537.0 & 430.3 &18.84 &-0.09 &\phantom{-}70 & 554.9 & 452.5 &18.34 &-0.11
& 120 & 584.5 & 469.6 &17.81 &-0.05 \cr
&&&&&&&&&\cr
 21 & 523.9 & 432.8 &17.08 &-0.04 &\phantom{-}71 & 515.3 & 452.6 &17.91 &-0.11
& 121 & 555.6 & 469.6 &18.05 &-0.14 \cr
 22 & 579.2 & 434.0 &18.39 &-0.09 &\phantom{-}72 & 452.9 & 452.8 &18.17 &-0.17
& 122 & 567.2 & 470.3 &17.54 &\phantom{-}0.02 \cr
 23 & 569.2 & 435.5 &16.51 &-0.01 &\phantom{-}73 & 572.1 & 452.9 &16.81 &-0.01
& 123 & 528.1 & 470.5 &18.41 &-0.34 \cr
 24 & 476.1 & 435.8 &17.03 &-0.01 &\phantom{-}74 & 494.1 & 453.8 &16.53
&\phantom{-}1.45 & 124 & 549.2 & 470.6 &18.45 &-0.16 \cr
 25 & 511.8 & 435.8 &17.62 &\phantom{-}0.08 &\phantom{-}75 & 541.1 & 454.9
&17.86 &-0.13 & 125 & 575.1 & 470.8 &17.11 &-0.01 \cr
 26 & 466.5 & 437.0 &18.06 &-0.05 &\phantom{-}76 & 491.9 & 455.3 &15.76
&\phantom{-}1.46 & 126 & 523.4 & 470.9 &17.77 &\phantom{-}0.12 \cr
 27 & 504.5 & 437.4 &18.75 &\phantom{-}0.31 &\phantom{-}77 & 556.4 & 455.6
&18.94 &-0.25 & 127 & 494.3 & 471.2 &13.69 &-0.14 \cr
 28 & 490.7 & 437.7 &18.15 &\phantom{-}0.07 &\phantom{-}78 & 515.5 & 456.9
&16.28 &-0.09 & 128 & 544.7 & 471.7 &17.58 &-0.10 \cr
 29 & 471.6 & 437.9 &15.92 &-0.06 &\phantom{-}79 & 590.4 & 457.0 &17.16 &-0.08
& 129 & 608.3 & 471.7 &18.34 &\phantom{-}0.07 \cr
 30 & 539.8 & 437.9 &16.81 &\phantom{-}0.02 &\phantom{-}80 & 492.2 & 457.1
&17.09 &\phantom{-}1.47 & 130 & 439.6 & 471.9 &17.60 &\phantom{-}0.07 \cr
&&&&&&&&&\cr
 31 & 480.3 & 439.0 &17.24 &\phantom{-}0.95 &\phantom{-}81 & 542.5 & 457.6
&15.83 &-0.12 & 131 & 552.8 & 472.2 &17.14 &-0.09 \cr
 32 & 502.0 & 439.2 &17.62 &-0.06 &\phantom{-}82 & 522.5 & 457.8 &16.92 &-0.08
& 132 & 445.7 & 472.6 &18.81 &-0.21 \cr
 33 & 572.7 & 439.2 &18.65 &\phantom{-}0.22 &\phantom{-}83 & 556.2 & 457.9
&17.64 &-0.05 & 133 & 530.5 & 473.3 &18.01 &-0.02 \cr
 34 & 531.5 & 439.2 &17.79 &-0.08 &\phantom{-}84 & 532.6 & 458.0 &18.18
&\phantom{-}0.18 & 134 & 473.9 & 473.6 &16.66 &-0.10 \cr
 35 & 486.7 & 440.0 &15.81 &\phantom{-}0.01 &\phantom{-}85 & 527.6 & 458.2
&14.32 &-0.16 & 135 & 513.8 & 473.7 &15.43 &\phantom{-}1.22 \cr
 36 & 512.0 & 440.3 &17.78 &-0.16 &\phantom{-}86 & 443.2 & 459.1 &18.44
&\phantom{-}0.00 & 136 & 524.2 & 474.4 &14.96 &\phantom{-}0.12 \cr
 37 & 581.7 & 441.1 &17.11 &-0.04 &\phantom{-}87 & 521.1 & 459.4 &15.43 &-0.14
& 137 & 450.7 & 474.7 &19.29 &\phantom{-}0.12 \cr
 38 & 589.4 & 441.5 &18.64 &\phantom{-}0.04 &\phantom{-}88 & 460.3 & 459.7
&15.06 &\phantom{-}1.02 & 138 & 613.5 & 474.8 &18.38 &\phantom{-}0.01 \cr
 39 & 572.5 & 441.5 &18.04 &-0.06 &\phantom{-}89 & 569.7 & 460.4 &18.61
&\phantom{-}0.03 & 139 & 459.8 & 475.1 &17.82 &\phantom{-}0.07 \cr
 40 & 464.3 & 441.7 &16.90 &\phantom{-}1.46 &\phantom{-}90 & 487.1 & 460.8
&17.82 &\phantom{-}0.09 & 140 & 568.6 & 475.6 &17.63 &-0.15 \cr
&&&&&&&&&\cr
 41 & 549.6 & 441.8 &18.11 &-0.04 &\phantom{-}91 & 539.8 & 461.2 &19.41 &-0.22
& 141 & 472.3 & 476.2 &17.20 &-0.11 \cr
 42 & 569.3 & 441.8 &19.04 &-0.07 &\phantom{-}92 & 524.3 & 461.5 &13.54 &-0.13
& 142 & 550.6 & 476.4 &16.81 &\phantom{-}0.00 \cr
 43 & 527.8 & 443.0 &15.39 &\phantom{-}1.28 &\phantom{-}93 & 492.7 & 461.7
&17.33 &-0.07 & 143 & 492.8 & 476.7 &17.07 &-0.07 \cr
 44 & 558.1 & 443.1 &17.75 &-0.12 &\phantom{-}94 & 534.0 & 461.9 &17.04 &-0.06
& 144 & 457.3 & 477.0 &17.50 &\phantom{-}1.11 \cr
 45 & 587.1 & 443.6 &16.20 &-0.04 &\phantom{-}95 & 557.8 & 462.6 &17.06 &-0.06
& 145 & 466.2 & 477.1 &17.60 &-0.09 \cr
 46 & 540.8 & 443.6 &18.43 &\phantom{-}0.20 &\phantom{-}96 & 469.1 & 462.8
&17.22 &-0.03 & 146 & 494.6 & 477.7 &17.64 &-0.16 \cr
 47 & 492.4 & 443.9 &17.94 &-0.06 &\phantom{-}97 & 523.6 & 463.2 &13.09 &-0.12
& 147 & 579.6 & 477.9 &17.76 &-0.02 \cr
 48 & 523.3 & 444.4 &18.11 &\phantom{-}0.00 &\phantom{-}98 & 495.5 & 463.2
&16.72 &\phantom{-}0.72 & 148 & 532.3 & 478.1 &18.17 &-0.08 \cr
 49 & 471.0 & 444.4 &19.15 &\phantom{-}0.06 &\phantom{-}99 & 451.6 & 463.4
&18.75 &\phantom{-}0.13 & 149 & 566.4 & 478.3 &13.74 &\phantom{-}0.24 \cr
 50 & 573.9 & 444.5 &17.68 &-0.05 &100 & 579.6 & 463.9 &17.67 &\phantom{-}0.10
& 150 & 588.6 & 478.5 &18.13 &\phantom{-}0.63 \cr
\splc
}
\endinsert
\vfill\dosupereject

\pageinsert
\fsmall{
\tabhl{\taonen}{BV Photometry}
\span\tcoli &
\span\tcol{.05} &
\span\tcol{.05} &
\span\tcol{.05} &
\span\tcoll{.05}{4.5}{3.5} &
\span\tcol{.1} &
\span\tcol{.05} &
\span\tcol{.05} &
\span\tcol{.05} &
\span\tcoll{.05}{4.5}{3.5} &&
\span\tcol{.1} \cr
\noalign{\vskip0.3cm}
\tablerule

&&&&&&&&&\cr
\hfil ID & X & Y & V & B--V & ID & X & Y & V & B--V & ID & X & Y & V & B--V \cr
&&&&&&&&&\cr
\tablerule
&&&&&&&&&\cr

 151 & 559.0 & 478.8 &17.77 &-0.03 & 201 & 535.1 & 492.3 &15.31
&\phantom{-}0.98 & 251 & 542.2 & 502.8 &14.76 &\phantom{-}0.12 \cr
 152 & 613.8 & 478.9 &14.85 &\phantom{-}0.50 & 202 & 586.1 & 492.4 &18.30
&-0.06 & 252 & 469.2 & 503.1 &17.36 &-0.04 \cr
 153 & 481.1 & 479.1 &18.48 &-0.17 & 203 & 540.5 & 492.7 &17.08
&\phantom{-}0.10 & 253 & 488.7 & 503.5 &18.32 &\phantom{-}0.04 \cr
 154 & 472.8 & 479.4 &16.73 &-0.00 & 204 & 469.1 & 493.1 &17.47
&\phantom{-}0.07 & 254 & 511.9 & 503.6 &14.72 &\phantom{-}0.13 \cr
 155 & 529.5 & 479.5 &17.41 &-0.12 & 205 & 572.0 & 493.3 &18.20
&\phantom{-}0.03 & 255 & 595.2 & 503.8 &18.15 &\phantom{-}0.14 \cr
 156 & 562.3 & 480.1 &18.37 &-0.06 & 206 & 495.2 & 493.5 &16.25 &-0.05 & 256 &
479.4 & 504.0 &18.11 &-0.04 \cr
 157 & 569.4 & 480.1 &17.32 &\phantom{-}0.04 & 207 & 567.3 & 493.7 &15.42
&-0.09 & 257 & 534.4 & 504.3 &16.86 &-0.08 \cr
 158 & 556.0 & 480.4 &17.33 &\phantom{-}0.00 & 208 & 463.9 & 493.8 &15.79
&-0.02 & 258 & 454.3 & 504.4 &15.65 &\phantom{-}1.24 \cr
 159 & 485.3 & 480.6 &18.58 &-0.24 & 209 & 554.9 & 493.8 &15.40
&\phantom{-}1.26 & 259 & 499.9 & 504.4 &13.56 &\phantom{-}0.65 \cr
 160 & 468.2 & 481.1 &16.56 &-0.05 & 210 & 432.3 & 494.2 &18.53 &-0.03 & 260 &
467.2 & 504.7 &17.38 &-0.05 \cr
&&&&&&&&&\cr
 161 & 591.7 & 481.2 &18.54 &-0.04 & 211 & 560.5 & 494.2 &18.63
&\phantom{-}0.32 & 261 & 494.5 & 505.0 &15.36 &\phantom{-}0.86 \cr
 162 & 574.4 & 481.4 &18.37 &-0.06 & 212 & 607.0 & 494.2 &18.31
&\phantom{-}0.01 & 262 & 484.4 & 505.4 &15.37 &\phantom{-}1.44 \cr
 163 & 581.5 & 481.8 &19.10 &\phantom{-}0.03 & 213 & 437.0 & 494.2 &14.85
&\phantom{-}0.19 & 263 & 580.5 & 505.6 &18.60 &\phantom{-}0.31 \cr
 164 & 587.7 & 482.1 &16.90 &\phantom{-}0.05 & 214 & 484.6 & 494.3 &17.68
&-0.02 & 264 & 477.8 & 506.2 &18.05 &\phantom{-}0.10 \cr
 165 & 502.4 & 482.7 &17.61 &-0.03 & 215 & 593.2 & 494.8 &18.78
&\phantom{-}0.05 & 265 & 447.1 & 506.6 &16.69 &-0.08 \cr
 166 & 536.7 & 482.9 &16.74 &-0.07 & 216 & 475.9 & 494.8 &17.99
&\phantom{-}0.03 & 266 & 552.3 & 507.5 &18.12 &-0.06 \cr
 167 & 508.6 & 483.1 &15.56 &\phantom{-}0.88 & 217 & 574.4 & 495.6 &18.96
&\phantom{-}0.33 & 267 & 584.0 & 507.7 &18.13 &\phantom{-}0.18 \cr
 168 & 462.0 & 483.2 &18.35 &-0.12 & 218 & 523.7 & 495.9 &16.01
&\phantom{-}0.09 & 268 & 574.3 & 507.9 &18.12 &\phantom{-}0.04 \cr
 169 & 442.6 & 483.7 &18.90 &-0.10 & 219 & 540.8 & 495.9 &17.22
&\phantom{-}0.06 & 269 & 568.8 & 507.9 &19.12 &-0.13 \cr
 170 & 598.9 & 484.2 &13.74 &-0.07 & 220 & 480.5 & 496.3 &16.53 &-0.02 & 270 &
558.2 & 508.4 &15.81 &\phantom{-}0.10 \cr
&&&&&&&&&\cr
 171 & 435.0 & 484.5 &19.30 &-0.32 & 221 & 547.4 & 496.4 &16.73 &-0.01 & 271 &
510.1 & 508.6 &15.86 &-0.24 \cr
 172 & 484.5 & 484.7 &18.18 &-0.03 & 222 & 508.1 & 496.4 &15.06
&\phantom{-}0.90 & 272 & 479.1 & 509.0 &17.12 &\phantom{-}0.03 \cr
 173 & 539.8 & 484.7 &14.26 &\phantom{-}0.78 & 223 & 598.0 & 496.7 &17.95
&-0.04 & 273 & 449.1 & 509.1 &18.48 &-0.03 \cr
 174 & 545.7 & 485.1 &17.19 &\phantom{-}0.11 & 224 & 453.2 & 496.7 &17.95
&-0.05 & 274 & 527.3 & 509.1 &16.19 &-0.17 \cr
 175 & 532.4 & 485.3 &15.21 &\phantom{-}1.40 & 225 & 491.2 & 497.0 &16.62
&-0.04 & 275 & 578.8 & 509.2 &18.41 &-0.02 \cr
 176 & 478.2 & 485.5 &17.64 &\phantom{-}0.18 & 226 & 467.5 & 497.2 &18.17
&-0.05 & 276 & 499.1 & 509.6 &16.85 &\phantom{-}0.09 \cr
 177 & 474.4 & 485.5 &14.94 &\phantom{-}0.27 & 227 & 483.9 & 497.4 &18.40
&\phantom{-}0.20 & 277 & 608.8 & 509.6 &17.20 &\phantom{-}0.07 \cr
 178 & 469.7 & 485.5 &18.30 &\phantom{-}0.14 & 228 & 504.5 & 497.6 &16.32
&-0.04 & 278 & 475.2 & 509.7 &17.50 &\phantom{-}0.00 \cr
 179 & 613.8 & 485.8 &18.76 &\phantom{-}0.09 & 229 & 487.2 & 497.7 &17.42
&\phantom{-}0.93 & 279 & 554.4 & 509.8 &17.41 &-0.07 \cr
 180 & 565.4 & 486.0 &18.93 &\phantom{-}0.28 & 230 & 559.2 & 498.3 &18.20
&\phantom{-}0.19 & 280 & 540.8 & 510.1 &17.94 &-0.22 \cr
&&&&&&&&&\cr
 181 & 526.8 & 486.0 &17.16 &-0.15 & 231 & 561.7 & 498.8 &18.42 &-0.03 & 281 &
491.3 & 510.2 &15.94 &-0.02 \cr
 182 & 535.9 & 486.0 &14.74 &\phantom{-}0.20 & 232 & 578.8 & 498.8 &17.23
&\phantom{-}0.00 & 282 & 432.2 & 510.6 &16.78 &-0.00 \cr
 183 & 440.6 & 486.3 &19.15 &-0.12 & 233 & 440.8 & 498.9 &18.68 &-0.02 & 283 &
604.6 & 511.1 &14.96 &\phantom{-}0.38 \cr
 184 & 595.7 & 486.4 &18.07 &-0.18 & 234 & 541.7 & 499.0 &17.90
&\phantom{-}0.01 & 284 & 528.5 & 511.2 &14.69 &\phantom{-}0.86 \cr
 185 & 448.4 & 487.3 &16.80 &\phantom{-}0.01 & 235 & 609.4 & 499.0 &17.93
&\phantom{-}0.00 & 285 & 425.5 & 511.3 &17.45 &-0.02 \cr
 186 & 463.1 & 487.6 &17.08 &\phantom{-}0.02 & 236 & 515.0 & 499.0 &15.86
&\phantom{-}1.01 & 286 & 587.2 & 511.5 &18.55 &\phantom{-}0.10 \cr
 187 & 458.5 & 487.8 &17.94 &\phantom{-}0.03 & 237 & 572.1 & 499.2 &18.42
&\phantom{-}0.07 & 287 & 440.1 & 511.6 &18.18 &\phantom{-}0.03 \cr
 188 & 561.9 & 487.9 &18.57 &\phantom{-}0.06 & 238 & 463.8 & 499.6 &16.29
&-0.13 & 288 & 485.5 & 511.6 &16.27 &-0.07 \cr
 189 & 468.8 & 488.0 &18.95 &-0.11 & 239 & 426.1 & 500.5 &16.96
&\phantom{-}0.03 & 289 & 567.5 & 511.7 &18.56 &-0.16 \cr
 190 & 504.4 & 488.3 &16.57 &-0.01 & 240 & 467.0 & 500.5 &17.52
&\phantom{-}0.06 & 290 & 512.7 & 511.9 &14.57 &\phantom{-}0.10 \cr
&&&&&&&&&\cr
 191 & 534.4 & 488.5 &16.33 &-0.04 & 241 & 445.7 & 501.3 &17.96 &-0.02 & 291 &
492.1 & 512.2 &14.87 &\phantom{-}0.17 \cr
 192 & 552.3 & 488.6 &18.20 &-0.07 & 242 & 437.6 & 501.4 &17.78
&\phantom{-}0.10 & 292 & 599.7 & 512.4 &16.83 &\phantom{-}0.12 \cr
 193 & 485.0 & 489.7 &18.23 &-0.13 & 243 & 464.9 & 501.5 &16.99
&\phantom{-}0.03 & 293 & 532.5 & 512.7 &14.62 &\phantom{-}0.15 \cr
 194 & 573.6 & 490.9 &16.53 &\phantom{-}0.06 & 244 & 481.2 & 501.5 &17.08
&\phantom{-}0.03 & 294 & 550.9 & 513.1 &17.03 &-0.28 \cr
 195 & 482.5 & 491.1 &17.91 &-0.24 & 245 & 497.1 & 501.8 &16.02 &-0.04 & 295 &
582.4 & 513.1 &18.50 &-0.02 \cr
 196 & 563.9 & 491.3 &18.11 &-0.01 & 246 & 430.1 & 502.4 &15.24
&\phantom{-}0.09 & 296 & 530.1 & 513.3 &14.85 &\phantom{-}0.18 \cr
 197 & 613.2 & 491.4 &18.48 &\phantom{-}0.17 & 247 & 475.6 & 502.5 &16.25
&-0.03 & 297 & 447.7 & 513.5 &17.99 &\phantom{-}0.06 \cr
 198 & 505.7 & 491.5 &15.30 &\phantom{-}0.13 & 248 & 601.9 & 502.5 &17.26
&-0.03 & 298 & 591.5 & 513.5 &18.53 &\phantom{-}0.09 \cr
 199 & 448.4 & 491.6 &18.77 &\phantom{-}0.25 & 249 & 535.8 & 502.6 &16.70
&-0.12 & 299 & 573.2 & 513.5 &18.63 &-0.15 \cr
 200 & 530.6 & 492.0 &17.28 &\phantom{-}0.01 & 250 & 574.4 & 502.7 &15.47
&\phantom{-}1.71 & 300 & 615.0 & 513.6 &17.80 &-0.03 \cr
\splc}
\endinsert
\vfill\dosupereject

\pageinsert
\fsmall{
\tabhl{\taonen}{BV Photometry}
\span\tcoli &
\span\tcol{.05} &
\span\tcol{.05} &
\span\tcol{.05} &
\span\tcoll{.05}{4.5}{3.5} &
\span\tcol{.1} &
\span\tcol{.05} &
\span\tcol{.05} &
\span\tcol{.05} &
\span\tcoll{.05}{4.5}{3.5} &&
\span\tcol{.1} \cr
\noalign{\vskip0.3cm}
\tablerule

&&&&&&&&&\cr
\hfil ID & X & Y & V & B--V & ID & X & Y & V & B--V & ID & X & Y & V & B--V \cr
&&&&&&&&&\cr
\tablerule
&&&&&&&&&\cr

 301 & 542.6 & 513.9 &16.48 &\phantom{-}0.01 & 351 & 560.8 & 526.2 &15.39
&\phantom{-}1.09 & 401 & 499.4 & 538.7 &16.38 &\phantom{-}0.03 \cr
 302 & 488.2 & 513.9 &14.76 &\phantom{-}1.21 & 352 & 538.0 & 526.3 &15.40
&-0.00 & 402 & 535.9 & 538.7 &15.33 &\phantom{-}0.01 \cr
 303 & 426.7 & 514.4 &18.95 &-0.20 & 353 & 575.7 & 526.4 &18.32
&\phantom{-}0.07 & 403 & 547.4 & 539.0 &17.33 &-0.16 \cr
 304 & 480.5 & 514.5 &15.46 &\phantom{-}1.53 & 354 & 471.4 & 526.5 &19.00
&\phantom{-}0.07 & 404 & 557.1 & 539.2 &17.01 &-0.10 \cr
 305 & 423.9 & 514.8 &17.40 &\phantom{-}0.04 & 355 & 540.7 & 526.7 &16.60
&-0.07 & 405 & 528.4 & 539.4 &17.88 &\phantom{-}0.22 \cr
 306 & 434.4 & 515.2 &18.19 &\phantom{-}0.23 & 356 & 520.2 & 526.9 &14.43
&\phantom{-}0.31 & 406 & 552.5 & 539.5 &16.86 &\phantom{-}0.01 \cr
 307 & 602.5 & 515.3 &17.80 &\phantom{-}0.16 & 357 & 509.0 & 527.4 &17.81
&\phantom{-}0.45 & 407 & 571.8 & 539.7 &18.47 &\phantom{-}0.08 \cr
 308 & 451.4 & 515.4 &17.97 &\phantom{-}0.16 & 358 & 552.0 & 528.2 &18.13
&\phantom{-}0.29 & 408 & 602.8 & 540.0 &18.01 &\phantom{-}0.20 \cr
 309 & 559.7 & 515.5 &16.57 &\phantom{-}0.16 & 359 & 466.5 & 528.2 &19.05
&-0.08 & 409 & 443.4 & 540.0 &18.02 &\phantom{-}0.06 \cr
 310 & 583.6 & 515.6 &17.95 &\phantom{-}0.00 & 360 & 580.5 & 528.2 &18.81
&\phantom{-}0.10 & 410 & 437.4 & 540.7 &18.71 &-0.04 \cr
&&&&&&&&&\cr
 311 & 576.4 & 515.6 &18.07 &\phantom{-}0.09 & 361 & 555.0 & 528.3 &17.76
&\phantom{-}0.17 & 411 & 524.8 & 540.7 &16.93 &\phantom{-}0.01 \cr
 312 & 462.4 & 515.9 &18.75 &-0.17 & 362 & 613.1 & 528.6 &15.44
&\phantom{-}2.08 & 412 & 612.0 & 540.9 &19.13 &\phantom{-}0.62 \cr
 313 & 550.7 & 516.2 &15.70 &-0.11 & 363 & 480.7 & 528.7 &17.84 &-0.09 & 413 &
483.6 & 541.0 &18.59 &-0.17 \cr
 314 & 540.9 & 516.6 &17.23 &\phantom{-}0.35 & 364 & 517.3 & 528.8 &16.63
&\phantom{-}0.03 & 414 & 435.2 & 541.2 &17.97 &\phantom{-}0.05 \cr
 315 & 497.6 & 517.0 &17.58 &-0.04 & 365 & 558.2 & 529.1 &16.83
&\phantom{-}0.12 & 415 & 443.5 & 542.2 &18.91 &\phantom{-}0.01 \cr
 316 & 558.6 & 517.0 &17.19 &\phantom{-}0.07 & 366 & 473.2 & 529.1 &16.95
&\phantom{-}0.04 & 416 & 495.7 & 542.2 &17.67 &\phantom{-}0.04 \cr
 317 & 580.7 & 517.0 &16.35 &\phantom{-}1.29 & 367 & 535.2 & 529.3 &14.53
&\phantom{-}0.17 & 417 & 588.5 & 542.5 &17.24 &-0.31 \cr
 318 & 530.2 & 517.7 &15.44 &\phantom{-}0.63 & 368 & 566.6 & 530.0 &17.73
&-0.09 & 418 & 530.9 & 543.0 &16.70 &\phantom{-}0.06 \cr
 319 & 542.5 & 518.4 &16.77 &\phantom{-}0.06 & 369 & 501.5 & 530.5 &15.68
&\phantom{-}1.64 & 419 & 511.7 & 543.3 &17.15 &-0.12 \cr
 320 & 477.0 & 518.5 &14.97 &\phantom{-}0.19 & 370 & 569.7 & 531.2 &15.64
&\phantom{-}1.46 & 420 & 556.2 & 543.5 &17.24 &\phantom{-}0.17 \cr
&&&&&&&&&\cr
 321 & 494.9 & 518.5 &17.78 &\phantom{-}0.09 & 371 & 467.4 & 531.2 &18.48
&\phantom{-}0.29 & 421 & 609.8 & 544.0 &19.23 &\phantom{-}0.12 \cr
 322 & 488.0 & 519.2 &15.34 &\phantom{-}1.24 & 372 & 589.8 & 531.5 &18.32
&\phantom{-}0.02 & 422 & 589.9 & 544.1 &14.84 &\phantom{-}1.75 \cr
 323 & 566.0 & 519.2 &17.21 &\phantom{-}0.13 & 373 & 498.7 & 532.1 &17.43
&-0.25 & 423 & 571.7 & 544.4 &18.28 &\phantom{-}0.14 \cr
 324 & 501.5 & 519.3 &15.58 &\phantom{-}0.30 & 374 & 561.8 & 532.2 &16.87
&\phantom{-}0.02 & 424 & 558.7 & 544.9 &15.36 &\phantom{-}1.55 \cr
 325 & 555.5 & 519.4 &17.29 &-0.24 & 375 & 508.3 & 532.3 &17.35
&\phantom{-}0.06 & 425 & 564.4 & 545.3 &18.16 &\phantom{-}0.20 \cr
 326 & 481.0 & 519.4 &18.11 &-0.03 & 376 & 539.9 & 532.5 &16.48
&\phantom{-}0.34 & 426 & 515.5 & 546.6 &15.60 &\phantom{-}1.34 \cr
 327 & 551.5 & 519.7 &17.53 &-0.49 & 377 & 514.5 & 533.0 &15.56
&\phantom{-}1.31 & 427 & 569.2 & 547.0 &16.03 &\phantom{-}0.10 \cr
 328 & 575.3 & 519.8 &17.61 &\phantom{-}0.09 & 378 & 528.9 & 533.2 &15.39
&\phantom{-}1.42 & 428 & 441.4 & 547.3 &17.96 &\phantom{-}0.02 \cr
 329 & 457.3 & 520.2 &17.62 &\phantom{-}0.10 & 379 & 517.2 & 533.3 &15.78
&\phantom{-}0.03 & 429 & 520.1 & 547.7 &17.21 &\phantom{-}0.00 \cr
 330 & 610.8 & 520.4 &14.99 &\phantom{-}0.31 & 380 & 556.8 & 533.5 &18.08
&\phantom{-}0.03 & 430 & 510.9 & 548.0 &17.16 &\phantom{-}0.02 \cr
&&&&&&&&&\cr
 331 & 543.1 & 520.6 &17.60 &-0.05 & 381 & 560.0 & 533.6 &16.84 &-0.05 & 431 &
559.6 & 548.0 &16.94 &-0.01 \cr
 332 & 484.4 & 520.7 &15.86 &\phantom{-}1.33 & 382 & 548.6 & 534.0 &14.81
&\phantom{-}0.19 & 432 & 471.2 & 548.1 &17.92 &\phantom{-}0.04 \cr
 333 & 498.2 & 520.7 &15.86 &\phantom{-}1.24 & 383 & 526.3 & 534.0 &16.70
&-0.16 & 433 & 488.0 & 548.2 &14.94 &\phantom{-}1.22 \cr
 334 & 433.3 & 520.9 &18.10 &-0.15 & 384 & 539.2 & 534.6 &15.27
&\phantom{-}0.15 & 434 & 434.0 & 548.5 &18.32 &-0.11 \cr
 335 & 505.2 & 521.6 &16.74 &\phantom{-}0.24 & 385 & 572.9 & 535.0 &15.79
&\phantom{-}0.06 & 435 & 475.9 & 548.5 &18.32 &-0.26 \cr
 336 & 467.7 & 521.8 &15.56 &\phantom{-}1.22 & 386 & 466.8 & 535.0 &18.52
&-0.03 & 436 & 499.9 & 548.8 &17.72 &-0.10 \cr
 337 & 443.7 & 522.2 &15.13 &\phantom{-}1.70 & 387 & 595.0 & 535.5 &16.90
&\phantom{-}0.01 & 437 & 528.8 & 549.9 &17.36 &\phantom{-}0.07 \cr
 338 & 478.7 & 522.5 &15.70 &\phantom{-}1.58 & 388 & 483.2 & 535.6 &15.64
&\phantom{-}1.27 & 438 & 586.5 & 550.2 &18.33 &\phantom{-}0.21 \cr
 339 & 564.5 & 522.9 &17.22 &-0.02 & 389 & 504.1 & 535.7 &18.06
&\phantom{-}0.03 & 439 & 547.5 & 550.2 &16.45 &\phantom{-}0.03 \cr
 340 & 585.2 & 522.9 &17.55 &\phantom{-}0.08 & 390 & 617.9 & 536.1 &19.36
&\phantom{-}0.21 & 440 & 451.8 & 550.7 &18.12 &\phantom{-}0.01 \cr
&&&&&&&&&\cr
 341 & 556.8 & 522.9 &14.89 &\phantom{-}0.24 & 391 & 570.5 & 536.2 &15.41
&\phantom{-}1.43 & 441 & 559.5 & 550.7 &15.71 &\phantom{-}1.28 \cr
 342 & 447.6 & 523.0 &18.07 &\phantom{-}0.11 & 392 & 510.1 & 536.2 &17.38
&\phantom{-}0.05 & 442 & 551.0 & 550.7 &17.89 &\phantom{-}0.15 \cr
 343 & 536.9 & 523.1 &17.87 &-0.12 & 393 & 543.0 & 536.4 &14.55
&\phantom{-}0.25 & 443 & 515.0 & 550.9 &17.16 &-0.09 \cr
 344 & 509.2 & 523.2 &16.30 &\phantom{-}0.04 & 394 & 560.2 & 536.5 &14.91
&\phantom{-}0.25 & 444 & 491.5 & 551.1 &18.06 &\phantom{-}0.08 \cr
 345 & 611.0 & 524.2 &17.34 &\phantom{-}0.42 & 395 & 491.0 & 536.7 &18.66
&\phantom{-}0.24 & 445 & 540.2 & 551.2 &17.22 &-0.72 \cr
 346 & 559.6 & 524.3 &14.53 &\phantom{-}0.11 & 396 & 556.7 & 536.9 &14.83
&\phantom{-}0.19 & 446 & 533.4 & 551.2 &17.36 &\phantom{-}0.18 \cr
 347 & 486.8 & 524.5 &17.24 &\phantom{-}0.00 & 397 & 435.0 & 537.1 &18.29
&\phantom{-}0.08 & 447 & 432.1 & 551.3 &18.52 &\phantom{-}0.15 \cr
 348 & 516.6 & 524.6 &15.97 &\phantom{-}0.13 & 398 & 427.1 & 538.0 &17.85
&-0.01 & 448 & 553.5 & 551.6 &17.01 &\phantom{-}0.02 \cr
 349 & 505.5 & 525.5 &15.59 &\phantom{-}1.05 & 399 & 600.6 & 538.3 &18.00
&\phantom{-}0.24 & 449 & 455.8 & 551.6 &17.14 &\phantom{-}0.00 \cr
 350 & 442.8 & 526.2 &18.51 &\phantom{-}0.04 & 400 & 524.0 & 538.3 &14.99
&\phantom{-}0.05 & 450 & 565.8 & 552.5 &17.79 &\phantom{-}0.02 \cr
\splc}
\endinsert
\vfill\dosupereject
\pageinsert
\fsmall{
\tabhl{\taonen}{BV Photometry}
\span\tcoli &
\span\tcol{.05} &
\span\tcol{.05} &
\span\tcol{.05} &
\span\tcoll{.05}{4.5}{3.5} &
\span\tcol{.1} &
\span\tcol{.05} &
\span\tcol{.05} &
\span\tcol{.05} &
\span\tcoll{.05}{4.5}{3.5} &&
\span\tcol{.1} \cr
\noalign{\vskip0.3cm}
\tablerule

&&&&&&&&&\cr
\hfil ID & X & Y & V & B--V & ID & X & Y & V & B--V & ID & X & Y & V & B--V \cr
&&&&&&&&&\cr
\tablerule
&&&&&&&&&\cr
 451 & 571.9 & 552.6 &18.65 &\phantom{-}0.00 & 501 & 497.5 & 564.2 &13.68
&\phantom{-}0.23 & 551 & 483.9 & 578.0 &18.48 &-0.05 \cr
 452 & 470.0 & 552.7 &18.05 &\phantom{-}0.26 & 502 & 484.0 & 565.6 &17.61
&-0.02 & 552 & 505.1 & 578.9 &17.34 &-0.03 \cr
 453 & 534.5 & 553.1 &17.35 &-0.13 & 503 & 471.4 & 566.3 &18.08
&\phantom{-}0.02 & 553 & 494.5 & 579.1 &16.59 &-0.00 \cr
 454 & 544.1 & 553.2 &17.38 &-0.14 & 504 & 527.4 & 566.6 &17.36
&\phantom{-}0.02 & 554 & 564.9 & 579.4 &19.23 &\phantom{-}0.03 \cr
 455 & 578.6 & 553.6 &17.81 &\phantom{-}0.16 & 505 & 607.7 & 566.6 &16.79
&\phantom{-}1.65 & 555 & 554.7 & 579.4 &16.89 &-0.01 \cr
 456 & 463.0 & 553.7 &18.96 &-0.21 & 506 & 562.8 & 566.6 &18.17 &-0.01 & 556 &
577.4 & 579.7 &18.72 &\phantom{-}0.29 \cr
 457 & 481.0 & 554.2 &16.42 &-0.12 & 507 & 595.6 & 567.3 &18.46
&\phantom{-}0.22 & 557 & 487.6 & 580.2 &18.59 &\phantom{-}0.11 \cr
 458 & 527.8 & 554.2 &15.96 &\phantom{-}1.34 & 508 & 574.6 & 567.4 &16.94
&\phantom{-}0.09 & 558 & 509.2 & 580.2 &18.63 &\phantom{-}0.07 \cr
 459 & 457.2 & 554.3 &16.76 &-0.05 & 509 & 585.8 & 567.6 &16.06
&\phantom{-}1.48 & 559 & 519.5 & 580.8 &16.89 &\phantom{-}0.08 \cr
 460 & 531.7 & 554.5 &17.02 &-0.10 & 510 & 441.2 & 567.8 &14.87
&\phantom{-}0.11 & 560 & 469.2 & 581.8 &15.32 &\phantom{-}0.14 \cr
&&&&&&&&&\cr
 461 & 545.5 & 554.8 &17.61 &-0.17 & 511 & 556.0 & 567.9 &18.87
&\phantom{-}0.41 & 561 & 556.2 & 582.1 &18.82 &-0.05 \cr
 462 & 559.3 & 555.2 &17.63 &\phantom{-}0.15 & 512 & 486.7 & 568.2 &18.29
&\phantom{-}0.01 & 562 & 575.5 & 582.2 &17.63 &\phantom{-}0.03 \cr
 463 & 460.4 & 555.6 &17.18 &\phantom{-}0.08 & 513 & 483.2 & 569.0 &18.13
&\phantom{-}0.06 & 563 & 462.8 & 582.2 &18.28 &-0.02 \cr
 464 & 485.8 & 555.6 &17.69 &\phantom{-}0.04 & 514 & 520.5 & 569.1 &14.68
&\phantom{-}0.30 & 564 & 581.4 & 582.3 &18.50 &\phantom{-}0.12 \cr
 465 & 448.6 & 555.7 &18.83 &\phantom{-}0.05 & 515 & 462.7 & 569.3 &19.03
&-0.25 & 565 & 505.0 & 583.3 &18.81 &\phantom{-}0.12 \cr
 466 & 435.8 & 555.8 &19.18 &\phantom{-}0.13 & 516 & 470.6 & 569.9 &18.41
&\phantom{-}0.07 & 566 & 511.0 & 584.0 &15.60 &\phantom{-}1.02 \cr
 467 & 529.0 & 555.9 &15.90 &\phantom{-}1.33 & 517 & 465.2 & 570.0 &17.34
&\phantom{-}0.18 & 567 & 473.9 & 584.6 &17.88 &\phantom{-}0.07 \cr
 468 & 611.1 & 556.1 &19.25 &\phantom{-}0.09 & 518 & 541.8 & 570.0 &17.73
&\phantom{-}0.10 & 568 & 561.5 & 585.4 &17.70 &\phantom{-}0.84 \cr
 469 & 509.6 & 556.4 &17.78 &-0.00 & 519 & 561.4 & 570.3 &18.12 &-0.06 & 569 &
460.4 & 585.7 &17.28 &\phantom{-}0.16 \cr
 470 & 478.5 & 556.5 &15.40 &\phantom{-}0.00 & 520 & 546.8 & 570.3 &17.76
&-0.04 & 570 & 523.5 & 585.8 &19.06 &-0.20 \cr
&&&&&&&&&\cr
 471 & 506.1 & 556.7 &15.46 &\phantom{-}1.63 & 521 & 479.0 & 570.3 &17.82
&-0.26 & 571 & 514.3 & 585.9 &18.76 &\phantom{-}0.30 \cr
 472 & 493.9 & 556.7 &17.72 &-0.08 & 522 & 598.3 & 570.3 &18.06
&\phantom{-}0.05 & 572 & 493.4 & 586.0 &19.07 &-0.34 \cr
 473 & 482.3 & 557.2 &18.22 &\phantom{-}0.07 & 523 & 514.1 & 570.5 &18.15
&\phantom{-}0.13 & 573 & 549.5 & 586.3 &18.42 &\phantom{-}0.14 \cr
 474 & 569.7 & 557.5 &17.92 &\phantom{-}0.11 & 524 & 584.9 & 571.2 &17.95
&\phantom{-}0.13 & 574 & 496.6 & 586.8 &17.85 &\phantom{-}0.12 \cr
 475 & 600.8 & 557.7 &19.09 &\phantom{-}0.13 & 525 & 553.0 & 571.4 &18.88
&\phantom{-}0.10 & 575 & 481.8 & 587.1 &18.55 &\phantom{-}0.08 \cr
 476 & 540.2 & 557.7 &17.84 &-0.08 & 526 & 481.9 & 571.5 &18.59
&\phantom{-}0.01 & 576 & 457.7 & 587.2 &16.07 &\phantom{-}1.41 \cr
 477 & 498.1 & 557.8 &17.77 &-0.08 & 527 & 451.8 & 572.0 &19.15
&\phantom{-}0.05 & 577 & 486.0 & 587.4 &17.25 &\phantom{-}0.23 \cr
 478 & 543.0 & 558.5 &17.82 &\phantom{-}0.09 & 528 & 496.8 & 572.2 &16.53
&\phantom{-}0.01 & 578 & 500.5 & 587.5 &18.73 &\phantom{-}0.14 \cr
 479 & 551.6 & 558.8 &17.92 &\phantom{-}0.05 & 529 & 475.6 & 572.2 &16.88
&-0.34 & 579 & 544.1 & 587.8 &18.06 &-0.08 \cr
 480 & 486.8 & 559.2 &16.39 &\phantom{-}1.43 & 530 & 557.1 & 572.2 &18.15
&-0.07 & 580 & 588.2 & 588.1 &19.18 &\phantom{-}0.04 \cr
&&&&&&&&&\cr
 481 & 526.7 & 559.8 &18.47 &\phantom{-}0.06 & 531 & 504.1 & 572.5 &18.76
&-0.22 & 581 & 490.8 & 588.1 &16.25 &\phantom{-}0.42 \cr
 482 & 582.0 & 560.1 &17.55 &\phantom{-}0.06 & 532 & 471.6 & 572.6 &16.54
&\phantom{-}0.08 & 582 & 502.7 & 588.9 &16.98 &\phantom{-}0.51 \cr
 483 & 474.2 & 560.2 &19.02 &\phantom{-}0.08 & 533 & 499.3 & 572.9 &16.15
&-0.08 & 583 & 549.2 & 589.2 &18.07 &\phantom{-}0.01 \cr
 484 & 523.2 & 560.6 &18.32 &-0.03 & 534 & 600.2 & 573.0 &18.59
&\phantom{-}0.30 & 584 & 505.7 & 589.8 &17.76 &-0.01 \cr
 485 & 485.2 & 560.6 &16.07 &\phantom{-}0.64 & 535 & 468.0 & 573.6 &18.01
&\phantom{-}0.24 & 585 & 529.0 & 589.9 &18.84 &\phantom{-}0.12 \cr
 486 & 500.5 & 560.9 &17.94 &\phantom{-}0.03 & 536 & 528.7 & 573.9 &18.52
&-0.13 & 586 & 478.2 & 590.0 &18.23 &\phantom{-}0.02 \cr
 487 & 536.0 & 561.0 &16.70 &-0.04 & 537 & 458.4 & 574.0 &18.87
&\phantom{-}0.18 & 587 & 583.6 & 590.0 &15.73 &\phantom{-}1.54 \cr
 488 & 446.7 & 562.2 &18.14 &\phantom{-}0.09 & 538 & 548.6 & 574.1 &19.24
&-0.10 & 588 & 522.7 & 590.6 &18.99 &\phantom{-}0.01 \cr
 489 & 559.7 & 562.3 &14.58 &\phantom{-}1.70 & 539 & 486.9 & 574.4 &19.00
&\phantom{-}0.15 & 589 & 514.0 & 590.9 &19.05 &\phantom{-}0.06 \cr
 490 & 512.3 & 562.6 &18.16 &\phantom{-}0.11 & 540 & 521.7 & 574.5 &18.58
&-0.53 & 590 & 539.9 & 591.2 &17.50 &-0.07 \cr
&&&&&&&&&\cr
 491 & 588.0 & 562.7 &17.81 &\phantom{-}0.02 & 541 & 535.9 & 574.6 &18.29
&-0.13 & 591 & 576.7 & 591.3 &19.01 &-0.09 \cr
 492 & 533.0 & 562.8 &17.98 &\phantom{-}0.08 & 542 & 493.2 & 574.9 &19.36
&\phantom{-}0.10 & 592 & 489.2 & 591.3 &15.50 &\phantom{-}1.11 \cr
 493 & 480.6 & 563.1 &18.37 &-0.03 & 543 & 454.5 & 575.4 &18.41
&\phantom{-}0.02 & 593 & 509.4 & 591.5 &18.40 &\phantom{-}0.08 \cr
 494 & 502.5 & 563.2 &18.03 &\phantom{-}0.08 & 544 & 477.4 & 575.5 &19.53
&-0.15 & 594 & 553.8 & 592.3 &17.77 &-0.04 \cr
 495 & 443.8 & 563.6 &17.23 &-0.07 & 545 & 573.3 & 575.5 &17.72
&\phantom{-}0.02 & 595 & 517.3 & 592.7 &18.32 &-0.01 \cr
 496 & 542.5 & 563.8 &16.15 &\phantom{-}1.48 & 546 & 553.6 & 575.6 &16.27
&\phantom{-}0.92 & 596 & 504.4 & 592.7 &18.14 &-0.01 \cr
 497 & 551.4 & 564.0 &15.18 &\phantom{-}0.39 & 547 & 543.0 & 576.6 &18.14
&-0.03 & 597 & 570.4 & 593.1 &14.85 &\phantom{-}0.17 \cr
 498 & 530.6 & 564.0 &16.07 &\phantom{-}0.02 & 548 & 497.2 & 577.2 &18.31
&\phantom{-}0.21 & 598 & 477.3 & 593.5 &16.78 &\phantom{-}0.07 \cr
 499 & 606.1 & 564.0 &17.81 &\phantom{-}0.31 & 549 & 546.6 & 577.6 &17.27
&\phantom{-}0.03 & 599 & 486.6 & 593.8 &16.75 &\phantom{-}1.70 \cr
 500 & 508.9 & 564.2 &17.17 &-0.02 & 550 & 463.3 & 577.9 &18.75
&\phantom{-}0.56 & 600 & 583.1 & 594.1 &15.29 &\phantom{-}1.14 \cr
\splc}
\endinsert
\vfill\dosupereject
\pageinsert
\fsmall{
\tabhl{\taonen}{BV Photometry}
\span\tcoli &
\span\tcol{.05} &
\span\tcol{.05} &
\span\tcol{.05} &
\span\tcoll{.05}{4.5}{3.5} &
\span\tcol{.1} &
\span\tcol{.05} &
\span\tcol{.05} &
\span\tcol{.05} &
\span\tcoll{.05}{4.5}{3.5} &&
\span\tcol{.1} \cr
\noalign{\vskip0.3cm}
\tablerule

&&&&&&&&&\cr
\hfil ID & X & Y & V & B--V & ID & X & Y & V & B--V & ID & X & Y & V & B--V \cr
&&&&&&&&&\cr
\tablerule
&&&&&&&&&\cr

 601 & 473.9 & 594.3 &17.55 &\phantom{-}0.09 & 616 & 566.0 & 598.7 &19.08
&-0.36 & 631 & 518.5 & 604.1 &19.30 &\phantom{-}0.49 \cr
 602 & 471.3 & 594.5 &17.59 &\phantom{-}0.01 & 617 & 479.5 & 599.1 &18.99
&\phantom{-}0.24 & 632 & 509.2 & 604.3 &16.61 &\phantom{-}0.05 \cr
 603 & 509.7 & 594.7 &17.49 &\phantom{-}0.12 & 618 & 539.2 & 599.1 &18.05
&\phantom{-}0.12 & 633 & 525.5 & 604.9 &18.08 &\phantom{-}0.03 \cr
 604 & 498.9 & 594.8 &17.65 &\phantom{-}0.28 & 619 & 529.0 & 599.4 &18.00
&\phantom{-}0.17 & 634 & 484.8 & 605.1 &16.65 &-0.04 \cr
 605 & 555.9 & 595.1 &17.52 &-0.03 & 620 & 542.3 & 599.6 &18.43
&\phantom{-}0.11 & 635 & 499.6 & 605.4 &14.69 &\phantom{-}0.39 \cr
 606 & 546.7 & 595.4 &18.71 &\phantom{-}0.04 & 621 & 559.2 & 599.6 &15.51
&\phantom{-}0.28 & 636 & 492.9 & 606.0 &17.60 &\phantom{-}0.05 \cr
 607 & 519.1 & 595.5 &18.01 &-0.05 & 622 & 473.7 & 599.7 &17.63
&\phantom{-}0.01 & 637 & 540.7 & 606.0 &18.26 &-0.04 \cr
 608 & 552.3 & 595.7 &17.67 &\phantom{-}0.12 & 623 & 514.1 & 600.1 &18.34
&\phantom{-}0.09 & 638 & 553.8 & 606.3 &17.18 &\phantom{-}0.01 \cr
 609 & 532.4 & 596.1 &18.60 &\phantom{-}0.03 & 624 & 485.8 & 601.7 &17.69
&\phantom{-}1.15 & 639 & 485.3 & 607.3 &17.06 &\phantom{-}0.12 \cr
 610 & 508.5 & 596.3 &17.93 &\phantom{-}0.09 & 625 & 496.6 & 602.0 &16.64
&\phantom{-}0.14 & 640 & 512.4 & 607.7 &18.30 &-0.06 \cr
&&&&&&&&&\cr
 611 & 564.2 & 597.0 &18.48 &-0.04 & 626 & 477.4 & 602.3 &16.82
&\phantom{-}0.66 & 641 & 504.8 & 610.1 &17.95 &\phantom{-}0.10 \cr
 612 & 492.4 & 597.1 &15.86 &\phantom{-}1.27 & 627 & 475.6 & 602.5 &17.15
&\phantom{-}0.10 & 642 & 528.8 & 612.0 &18.30 &\phantom{-}1.04 \cr
 613 & 483.0 & 597.1 &18.22 &\phantom{-}0.05 & 628 & 491.8 & 602.7 &18.27
&-0.14 & 643 & 534.9 & 614.7 &18.06 &-0.10 \cr
 614 & 486.4 & 597.9 &18.39 &\phantom{-}0.54 & 629 & 482.0 & 603.0 &18.59
&\phantom{-}0.16 \cr
 615 & 500.9 & 598.5 &19.30 &-0.01 & 630 & 494.3 & 603.9 &17.67
&\phantom{-}0.02 \cr
\spl}
\endinsert

\midinsert
\tabhl{\taone}{Surface Photometry}
\span\tcoli & \span\tcoll{.3}{9}{4} && \span\tcol{.3} \cr
\topperl

\hfil \Rc2 & L$_B$ & \Rc2 & L$_V$ \cr
\hfil (pc) & (L$_B$\sol\ pc$^{-2}$) & (pc) & (L$_V$\sol\ pc$^{-2}$) \cr
\tablerule

  ~0.3 &  26246.0 $\pm$ 3517.0 & ~0.3 &  14938.0 $\pm$ 1958.0 \cr
  ~0.5 &  25094.0 $\pm$ 3993.0 & ~0.5 &  ~9719.0 $\pm$ 1844.0 \cr
  ~0.7 &  24901.0 $\pm$ 3821.0 & ~0.7 &  10536.0 $\pm$ 2155.0 \cr
  ~0.8 &  31223.0 $\pm$ 5737.0 & ~0.8 &  12438.0 $\pm$ 3273.0 \cr
  ~1.1 &  26021.0 $\pm$ 6693.0 & ~1.0 &  11899.0 $\pm$ 3994.0 \cr
  ~1.3 &  22898.0 $\pm$ 4908.0 & ~1.3 &  11771.0 $\pm$ 2702.0 \cr
  ~1.7 &  19512.0 $\pm$ 4085.0 & ~1.7 &  10236.0 $\pm$ 2041.0 \cr
  ~2.1 &  14890.0 $\pm$ 3159.0 & ~2.1 &  ~6037.0 $\pm$ 1010.0 \cr
  ~2.6 &  11276.0 $\pm$ 1308.0 & ~2.7 &  ~5831.0 $\pm$ ~988.0 \cr
  ~3.3 &  10828.0 $\pm$ 1290.0 & ~3.3 &  ~5357.0 $\pm$ ~554.0 \cr
  ~4.2 &  ~6910.0 $\pm$ 1417.0 & ~4.2 &  ~3517.0 $\pm$ ~733.0 \cr
  ~5.3 &  ~3889.0 $\pm$ ~938.0 & ~5.3 &  ~2098.0 $\pm$ ~443.0 \cr
  ~6.6 &  ~3198.0 $\pm$ ~435.0 & ~6.6 &  ~1625.0 $\pm$ ~261.0 \cr
  ~8.4 &  ~2919.0 $\pm$ ~429.0 & ~8.4 &  ~1423.0 $\pm$ ~199.0 \cr
  10.5 &  ~1044.0 $\pm$ ~210.0 & 10.5 &  ~~510.0 $\pm$ ~~88.0 \cr
  13.2 &  ~~944.0 $\pm$ ~201.0 & 13.2 &  ~~423.0 $\pm$ ~101.0 \cr
  16.7 &  ~~414.0 $\pm$ ~144.0 & 16.7 &  ~~233.0 $\pm$ ~~68.0 \cr
  21.0 &  ~~360.0 $\pm$ ~~63.0 & 21.0 &  ~~170.0 $\pm$ ~~29.0 \cr
  26.4 &  ~~140.0 $\pm$ ~~44.0 & 26.4 &  ~~~64.0 $\pm$ ~~23.0 \cr
  33.1 &  ~~~92.0 $\pm$ ~~39.0 & 33.1 &  ~~~46.0 $\pm$ ~~20.0 \cr
  40.5 &  ~~~22.0 $\pm$ ~~28.0 & 40.5 &  ~~~~8.0 $\pm$ ~~17.0 \cr
\spl
\endinsert

\midinsert
\tabhl{\taone}{King-Michie - B Band Fitted Parameters}
\span\tcolil{12}{4} & \hspa{.3} # & #\hfil & \span\tcol{.3} & \span\tcol{.3} &
#\hfil & \span\tcol{.3} & \span\tcoll{.3}{12}{4} && \span\tcol{.3} \cr
\topperl

 & & \multispan5 Photometry & & \multispan3 Velocities \cr
\tablerule
\hfil r$_a$ & \multispan2 W$_\circ$ & r$_s$ & \multispan2 c & $\chi_\nu^2$ &
P($> \chi_\nu^2$) & v$_s$ & $\zeta^2$ & P($> \zeta^2$) \cr
\hfil (r$_s$) & & & (pc) & & & ($\nu=18$) & & (km s$^{-1}$) & & \cr
\tablerule

 ISO & 7.\rlap3 & $\pm$ 0.30 & 2.6 $\pm$ 0.20 & ~42\rlap. & $\pm$ ~8. & 0.68 &
0.81 & 2.2 $\pm 0.4$ & 37.61 & 0.21 \cr
 ~20 & 7.\rlap3 & $\pm$ 0.30 & 2.6 $\pm$ 0.20 & ~46\rlap. & $\pmm{+12.}{!-9.}$
& 0.69 & 0.80 & 2.2 $\pm 0.4$ & 37.66 & 0.21  \cr
 ~10 & 7.\rlap3 & $\pm$ 0.35 & 2.6 $\pm$ 0.15 & ~67\rlap. & $\pmm{+45.}{-25.}$
& 0.72 & 0.76 & 2.2 $\pm 0.4$ & 37.80 & 0.20  \cr
 ~~5 & 6.\rlap8 & $\pmm{+0.10}{-0.20}$ & 3.1 $\pm$ 0.15 & 125\rlap. & $\pm$ 50.
& 0.86 & 0.60 & 2.2 $\pm 0.5$ & 38.40 & 0.15  \cr
 ~~3 & 5.\rlap9 & $\pmm{+0.05}{-0.10}$ & 3.5 $\pm$ 0.20 & 122\rlap. & $\pm$ 45.
& 1.17 & 0.25 & 2.3 $\pm 0.5$ & 39.11 & 0.10  \cr
\spl
\endinsert

\midinsert
\tabhl{Table \taone}{King-Michie - V Band Fitted Parameters}
\span\tcolil{12}{4} & \hspa{.3} # & #\hfil & \span\tcol{.3} & \span\tcol{.3} &
#\hfil & \span\tcol{.3} & \span\tcoll{.3}{12}{4} && \span\tcol{.3} \cr
\topperl

 & & \multispan5 Photometry & & \multispan3 Velocities \cr
\tablerule
\hfil r$_a$ & \multispan2 W$_\circ$ & r$_s$ & \multispan2 c & $\chi_\nu^2$ &
P($> \chi_\nu^2$) & v$_s$ & $\zeta^2$ & P($> \zeta^2$) \cr
\hfil (r$_s$) & & & (pc) & & & ($\nu=18$) & & (km s$^{-1}$) & & \cr
\tablerule

 ISO & 7.\rlap1 & $\pm$ 0.30 & 2.7 $\pm$ 0.20 & ~37\rlap. & $\pm$ ~8. & 0.91 &
0.51 & 2.2 $\pm 0.4$ & 37.68 & 0.21 \cr
 ~20 & 7.\rlap1 & $\pm$ 0.30 & 2.8 $\pm$ 0.20 & ~39\rlap. & $\pmm{+11.}{!-9.}$
& 0.91 & 0.51 & 2.2 $\pm 0.4$ & 37.72 & 0.20  \cr
 ~10 & 7.\rlap1 & $\pm$ 0.35 & 2.8 $\pm$ 0.20 & ~51\rlap. & $\pmm{+33.}{-16.}$
& 0.92 & 0.50 & 2.2 $\pm 0.4$ & 37.84 & 0.19 \cr
 ~~5 & 6.\rlap8 & $\pmm{+0.10}{-0.20}$ & 3.2 $\pm$ 0.20 & 116\rlap. & $\pm$ 47.
& 0.98 & 0.43 & 2.2 $\pm 0.5$ & 38.31 & 0.15  \cr
 ~~3 & 5.\rlap9 & $\pmm{+0.05}{-0.10}$ & 3.7 $\pm$ 0.20 & 122\rlap. & $\pm$ 44.
& 1.16 & 0.25 & 2.3 $\pm 0.5$ & 38.97 & 0.11  \cr
\spl
\endinsert

\midinsert
\tabh{\taone}{King-Michie - B Band Derived Parameters}
\span\tcoli && \span\tcol{.3} \cr
\topper

\hfil $r_a$ & $\rho_{KB\circ}$ & L$_B$ & Mass & M/L$_B$ \cr
\hfil ($r_s$) & (L$_B$\sol\ pc$^{-3}$) & (10$^6$L$_B$\sol) & (10$^4$M$_\odot$)
& (M$_\odot$/L$_B$\sol) \cr
\spacer

 ISO & 5500. $\pm$ 700. & 2.60 $\pm$ 0.20 & 5.7 $\pm 2.3$ & 0.02 $\pm 0.01$ \cr
 ~20 & 5500. $\pm$ 700. & 2.64 $\pm$ 0.20 & 5.7 $\pm 2.3$ & 0.02 $\pm 0.01$ \cr
 ~10 & 5400. $\pm$ 700. & 2.75 $\pm$ 0.30 & 5.9 $\pm 2.4$ & 0.02 $\pm 0.01$ \cr
 ~~5 & 4200. $\pm$ 550. & 2.75 $\pm$ 0.20 & 5.8 $\pm 2.3$ & 0.02 $\pm 0.01$ \cr
 ~~3 & 3700. $\pm$ 500. & 2.62 $\pm$ 0.20 & 5.4 $\pm 2.2$ & 0.02 $\pm 0.01$ \cr
\sp
\endinsert

\midinsert
\tabh{\taone}{King-Michie - V Band Derived Parameters}
\span\tcoli && \span\tcol{.3} \cr
\topper

\hfil $r_a$ & $\rho_{KV\circ}$ & L$_V$ & Mass & M/L$_V$ \cr
\hfil ($r_s$) & (L$_V$\sol\ $pc^{-3}$) & (10$^6$L$_V$\sol) & (10$^4$M\sol) &
(M\sol/L$_V$\sol) \cr
\spacer

 ISO & 2300. $\pm$ 350. & 1.26 $\pm$ 0.10 & 5.7 $\pm 2.3$ & 0.05 $\pm 0.02$ \cr
 ~20 & 2300. $\pm$ 350. & 1.27 $\pm$ 0.10 & 5.7 $\pm 2.3$ & 0.04 $\pm 0.02$ \cr
 ~10 & 2300. $\pm$ 300. & 1.31 $\pm$ 0.15 & 5.8 $\pm 2.3$ & 0.04 $\pm 0.02$ \cr
 ~~5 & 1900. $\pm$ 250. & 1.37 $\pm$ 0.10 & 5.9 $\pm 2.4$ & 0.04 $\pm 0.02$ \cr
 ~~3 & 1600. $\pm$ 200. & 1.32 $\pm$ 0.10 & 5.6 $\pm 2.2$ & 0.04 $\pm 0.02$ \cr
\sp
\endinsert

\midinsert
\tabh{\taone}{Velocity Standards}
\span\tcoli && \span\tcol{.3} \cr
\topper

\hfil HJD & v$_x$ & $\left<\hbox{v}_x\right>$ & $\chi^2_\nu$ \cr
\hfil (--2448000) & (km s$^{-1}$) & (km s$^{-1}$) & \cr
\spacer
\multispan4 \underbar{Local} \cr
&&&\cr
 304.5946 & 34.5 $\pm$ 0.3 & 34.2 $\pm$ 0.1 & 1.12 \cr
 305.5481 & 34.3 $\pm$ 0.3 &   & \cr
 306.5467 & 33.7 $\pm$ 0.3 &   & \cr
 307.5481 & 34.4 $\pm$ 0.2 &   & \cr
 308.5432 & 34.0 $\pm$ 0.3 &   & \cr
&&&\cr
 605.7090 & 36.0 $\pm$ 0.6 & 35.7 $\pm$ 0.3 & 0.72 \cr
 606.6605 & 36.0 $\pm$ 0.6 &   & \cr
 607.6073 & 35.0 $\pm$ 0.6 &   & \cr
 608.6622 & 35.7 $\pm$ 0.6 &   & \cr
&&&\cr
\multispan4 \underbar{HD~23214} \cr
&&&\cr
 305.5204 & -2.6 $\pm$ 0.4 & -2.9 $\pm$ 0.2 & 0.26 \cr
 306.5211 & -2.9 $\pm$ 0.4 &   & \cr
 307.5189 & -3.0 $\pm$ 0.4 &   & \cr
 308.5182 & -3.1 $\pm$ 0.4 &   & \cr
&&&\cr
 605.5891 & -3.1 $\pm$ 0.5 & -2.8 $\pm$ 0.2 & 0.32 \cr
 605.6899 & -2.8 $\pm$ 0.6 &   & \cr
 606.5868 & -2.9 $\pm$ 0.6 &   & \cr
 607.5920 & -2.4 $\pm$ 0.5 &   & \cr
 608.5895 & -2.7 $\pm$ 0.5 &   & \cr
\sp
\endinsert

\midinsert
\tabh{\taone}{Radial Velocities}
\span\tcoli && \span\tcol{.3} \cr
\topper

\hfil ID & \Rc2 & $\Theta$ & v$_x$ & $\left<\hbox{v}_x\right>$ & HJD & V & B--V
\cr
\hfil  & (pc) & ($^\circ$) & (km s$^{-1}$) & (km s$^{-1}$) & (--2448000) &
(mag) & (mag) \cr
\spacer

 ~RV1 &  0.0 &  ~18.5 &  252.7 $\pm$  2.1 &  252.7 $\pm$  2.1 & 608.6797 &
15.74 &   0.58 \cr
 ~RV2 &  0.5 &  166.7 &  248.3 $\pm$  1.3 &  248.3 $\pm$  1.3 & 608.6768 &
14.89 &   0.20 \cr
 ~RV3 &  0.9 &  224.6 &  248.4 $\pm$  1.9 &  248.4 $\pm$  1.9 & 304.5439 &
14.69 &   0.86 \cr
 ~RV4 &  1.0 &  163.4 &  251.8 $\pm$  2.6 &  251.8 $\pm$  2.6 & 608.6729 &
15.44 &   0.67 \cr
 ~RV5 &  1.1 &  325.4 &  248.4 $\pm$  4.9 &  248.4 $\pm$  4.9 & 303.5889 &
14.04 &   0.06 \cr
 ~RV6 &  1.3 &  200.3 &  249.8 $\pm$  3.3 &  249.8 $\pm$  3.3 & 303.6309 &
14.68 &   0.14 \cr
 ~RV7 &  1.5 &  ~71.1 &  252.2 $\pm$  1.9 &  252.2 $\pm$  1.9 & 303.5547 &
14.30 &  \llap{-}0.19 \cr
 ~RV8 &  1.7 &  349.2 &  260.2 $\pm$  4.3 &  260.2 $\pm$  4.3 & 304.5420 &
14.54 &   0.14 \cr
 ~RV9 &  1.7 &  102.0 &  254.3 $\pm$  1.7 &  252.3 $\pm$  1.1 & 607.6504 &
14.23 &   0.68 \cr
    &      &        &  251.0 $\pm$  1.4 &                   & 608.6973 &   &
\cr
 RV10 &  2.4 &  321.1 &  257.4 $\pm$  3.0 &  257.4 $\pm$  3.0 & 306.5625 &
14.72 &   0.13 \cr
 RV11 &  2.7 &  134.0 &  276.0 $\pm$  2.8 &  276.0 $\pm$  2.8 & 303.5850 &
14.53 &   0.18 \cr
 RV12 &  2.8 &  110.9 &  250.3 $\pm$  1.1 &  250.1 $\pm$  0.8 & 607.6543 &
15.39 &   1.40 \cr
    &      &        &  249.9 $\pm$  1.2 &                   & 608.7061 &  &
\cr
 RV13 &  2.8 &  340.9 &  254.6 $\pm$  1.3 &  254.6 $\pm$  1.3 & 608.6816 &
15.14 &   0.60 \cr
 RV14 &  3.0 &  ~67.9 &  253.1 $\pm$  2.2 &  253.1 $\pm$  2.2 & 608.6943 &
15.56 &   1.30 \cr
 RV15 &  3.0 &  236.9 &  246.7 $\pm$  1.6 &  246.7 $\pm$  1.6 & 607.6572 &
15.37 &   0.04 \cr
 RV16 &  3.1 &  ~76.5 &  248.5 $\pm$  2.2 &  248.5 $\pm$  2.2 & 304.5586 &
14.74 &   0.42 \cr
 RV17 &  3.3 &  218.8 &  243.4 $\pm$  3.1 &  243.4 $\pm$  3.1 & 304.5479 &
14.73 &   0.13 \cr
 RV18 &  3.6 &  315.3 &  234.4 $\pm$  2.4 &  234.4 $\pm$  2.4 & 306.5605 &
15.05 &   0.93 \cr
 RV19 &  3.8 &  341.1 &  272.7 $\pm$  1.3 &  260.7 $\pm$  1.0 & 303.5586 &
13.54 &   0.65 \cr
 RV19 &      &        &  240.9 $\pm$  1.7 &                   & 607.6416 &   &
  \cr
 RV20 &  4.0 &  ~39.9 &  253.6 $\pm$  1.3 &  253.2 $\pm$  0.9 & 608.6885 &
15.70 &   1.56 \cr
    &      &        &  252.9 $\pm$  1.2 &                   & 605.7393 &   &
\cr
 RV21 &  4.1 &  ~84.4 &  256.7 $\pm$  1.4 &  256.7 $\pm$  1.4 & 605.7344 &
15.19 &   1.62 \cr
 RV22 &  4.2 &  137.2 &  247.7 $\pm$  1.7 &  247.7 $\pm$  1.7 & 306.5576 &
14.55 &   0.23 \cr
 RV23 &  4.6 &  147.9 &  258.7 $\pm$  3.4 &  258.7 $\pm$  3.4 & 303.5752 &
14.82 &   0.16 \cr
 RV24 &  4.6 &  252.5 &  253.0 $\pm$  2.0 &  250.1 $\pm$  1.0 & 304.5547 &
14.71 &   0.21 \cr
    &      &        &  249.1 $\pm$  1.2 &                   & 608.6689 &   &
\cr
 RV25 &  4.7 &  ~~0.1 &  242.4 $\pm$  2.9 &  242.4 $\pm$  2.9 & 307.5625 &
14.85 &   0.15 \cr
\spc
\endinsert

\midinsert
\tabh{\taonen}{Radial Velocities}
\span\tcoli && \span\tcol{.3} \cr
\topperc

\hfil ID & \Rc2 & $\Theta$ & V$_x$ & $\left<\hbox{v}_x\right>$ & HJD & V & B--V
\cr
\hfil  & (pc) & ($^\circ$) & (km s$^{-1}$) & (km s$^{-1}$) & (--2448000) &
(mag) & (mag) \cr
\spacer

 RV26 &  5.0 &  247.4 &  242.3 $\pm$  2.9 &  242.3 $\pm$  2.9 & 304.5518 &
14.25 &   0.77 \cr
 RV27 &  5.1 &  ~57.9 &  248.3 $\pm$  1.6 &  248.3 $\pm$  1.6 & 608.6924 &
15.57 &   0.41 \cr
 RV28 &  5.2 &  300.6 &  255.7 $\pm$  2.1 &  255.7 $\pm$  2.1 & 608.7168 &
15.55 &   0.86 \cr
 RV29 &  5.2 &  ~~3.4 &  250.5 $\pm$  1.0 &  249.8 $\pm$  0.7 & 608.6855 &
14.77 &   1.18 \cr
    &      &        &  249.2 $\pm$  0.9 &                   & 307.5654 &  &
\cr
 RV30 &  5.5 &  219.5 &  253.5 $\pm$  1.5 &  253.5 $\pm$  1.5 & 608.7139 &
15.40 &   1.23 \cr
 RV31 &  6.0 &  351.4 &  252.2 $\pm$  1.7 &  252.2 $\pm$  1.7 & 606.6895 &
15.39 &   1.41 \cr
 RV32 &  6.3 &  288.7 &  250.7 $\pm$  2.0 &  250.7 $\pm$  2.0 & 606.6836 &
15.42 &   1.19 \cr
 RV33 &  6.7 &  ~14.6 &  251.9 $\pm$  1.3 &  251.9 $\pm$  1.3 & 605.7305 &
15.70 &   1.53 \cr
 RV34 &  6.9 &  ~~9.4 &  251.9 $\pm$  2.0 &  251.9 $\pm$  2.0 & 605.7266 &
14.96 &   0.19 \cr
 RV35 &  7.2 &  ~48.0 &  249.7 $\pm$  1.5 &  249.7 $\pm$  1.5 & 607.6465 &
14.94 &   1.19 \cr
 RV36 &  7.7 &  198.9 &  250.9 $\pm$  1.1 &  250.9 $\pm$  1.1 & 605.7227 &
15.48 &   1.64 \cr
 RV37 &  7.8 &  309.1 &  277.3 $\pm$  3.9 &  277.3 $\pm$  3.9 & 303.5625 &
13.67 &  \llap{-}0.14 \cr
 RV38 &  8.0 &  ~91.7 &  241.7 $\pm$  1.6 &  241.7 $\pm$  1.6 & 303.6201 &
14.67 &   0.29 \cr
 RV39 &  8.2 &  ~67.0 &  233.5 $\pm$  1.1 &  233.2 $\pm$  1.0 & 303.6152 &
13.67 &   0.23 \cr
    &      &        &  232.2 $\pm$  2.0 &                   & 607.6436 &   &
\cr
 RV40 &  8.3 &  226.0 &  239.6 $\pm$  3.8 &  239.6 $\pm$  3.8 & 303.5508 &
13.73 &   0.22 \cr
 RV41 &  8.3 &  ~12.0 &  252.5 $\pm$  1.3 &  252.5 $\pm$  1.3 & 608.7246 &
15.57 &   1.25 \cr
 RV42 &  8.3 &  301.6 &  252.5 $\pm$  2.0 &  252.5 $\pm$  2.0 & 308.5596 &
14.64 &   0.29 \cr
 RV43 &  8.7 &  132.5 &  249.1 $\pm$  1.1 &  249.1 $\pm$  1.1 & 607.6270 &
14.58 &   1.67 \cr
 RV44 & 10.3 &  356.5 &  250.1 $\pm$  1.7 &  250.1 $\pm$  1.7 & 608.7207 &
15.65 &   1.21 \cr
 RV45 & 10.6 &  161.5 &  250.7 $\pm$  1.0 &  250.7 $\pm$  1.0 & 605.7178 &
14.85 &   1.73 \cr
 RV46 & 11.8 &  ~10.2 &  252.4 $\pm$  1.5 &  252.4 $\pm$  1.5 & 606.6943 &
15.13 &   1.66 \cr
 RV47 & 13.7 &  ~80.2 &  249.6 $\pm$  1.5 &  249.6 $\pm$  1.5 & 305.5654 &
14.68 &   0.39 \cr
 RV48 & 16.6 &  304.7 &  252.0 $\pm$  1.2 &  252.0 $\pm$  1.2 & 607.6318 &
15.19 &   1.55 \cr
 RV49 & 18.3 &  120.6 &  288.6 $\pm$  1.1 &  286.3 $\pm$  0.7 & 607.6240 &
13.66 &   1.82 \cr
    &      &        &  284.4 $\pm$  1.0 &                   & 308.5518 &   &
\cr
 RV50 & 18.7 &  ~67.2 &  253.3 $\pm$  1.9 &  253.3 $\pm$  1.9 & 305.5586 &
14.54 &   0.49 \cr
 RV51 & 19.4 &  332.6 &  246.4 $\pm$  2.6 &  246.4 $\pm$  2.6 & 607.6377 &
15.08 &   1.41 \cr
 RV52 & 35.5 &  ~81.7 &  254.9 $\pm$  1.0 &  254.9 $\pm$  1.0 & 307.5576 &
14.59 &   1.71 \cr
\sp
\endinsert

\midinsert
\tabh{\taone}{Velocity Residuals}
\span\tcoli && \span\tcol{.3} \cr
\topper

\hfil Iteration & $v_s$ & $\delta_{max}$ & P($>\delta_{max}$) & Star \cr
\hfil  & (\kms) & & \cr
\spacer

1 & 4.8 & 4.35 & 0.00 & 39 \cr
2 & 2.8 & 3.39 & 0.01 & 38 \cr
3 & 2.1 & 2.49 & 0.39 & 52 \cr
\sp
\endinsert

\midinsert
\tabh{\taone}{Rotating Ellipsoidal Models}
\span\tcoli && \span\tcol{.3} \cr
\topper

\hfil $\epsilon$ & Mass & M/L$_B$ & M/L$_V$ & $\zeta^2$ & P($>\zeta^2$) & A \cr
\hfil & ($10^4$ M\sol) & (M\sol/L$_B$\sol) & (M\sol/L$_V$\sol) & & & (\kms) \cr
\spacer

 0.1 & 5.5 $\pm$ 0.2 & 0.02 $\pm$ 0.01 & 0.04 $\pm$ 0.02 & 37.47 & 0.34 & 0.9
\cr
 0.2 & 5.6 $\pm$ 0.2 & 0.02 $\pm$ 0.01 & 0.04 $\pm$ 0.02 & 38.76 & 0.37 & 0.5
\cr
 0.3 & 5.5 $\pm$ 0.2 & 0.02 $\pm$ 0.01 & 0.04 $\pm$ 0.02 & 40.58 & 0.20 & 0.3
\cr
\sp
\endinsert

\midinsert
\tabhl{\taone}{Mass Functions}
\span\tcoli & \span\tcol{.3} & \span\tcol{.3} & \span\tcoll{.3}{12}{4} &&
\span\tcol{.3} \cr
\topperl
\multispan6 M/L = \hfil\hfil\hskip0.03truecm ~~~~~~~~~0.02 $\pm$ 0.01
M\sol/L$_B$\sol\hfil \vbs{12}{4}\hfil ~~~0.05 $\pm$ 0.02 M\sol/L$_V$\sol
\hfil\cr
\tablerule
\hfil $m_l$ & $m_d$ & $x$ & $<m>$ & $x$ & $<m>$ \cr
\hfil (M\sol) & (M\sol) & & (M\sol) & & (M\sol) \cr
\tablerule

 0.05 & 0.0 & $0.29\pmm{+0.3}{-0.8}$ & 0.75 & $0.71\pmm{+0.2}{-0.4}$ & 0.37 \cr
 0.05 & 0.3 & $0.31\pmm{+0.3}{-0.8}$ & 0.83 & $0.79\pmm{+0.2}{-0.4}$ & 0.50 \cr
 0.05 & 0.5 & $0.32\pmm{+0.3}{-0.9}$ & 0.87 & $0.85\pmm{+0.3}{-0.5}$ & 0.57 \cr
 0.05 & 1.0 & $0.38\pmm{+0.5}{-1.0}$ & 0.93 & $1.04\pmm{+0.4}{-0.6}$ & 0.69 \cr
 0.05 & 1.5 & $0.46\pmm{+0.5}{-1.2}$ & 0.97 & $1.28\pmm{+0.5}{-0.8}$ & 0.77 \cr
\tablerule
 0.1~ & 0.0 & $0.32\pmm{+0.3}{-0.8}$ & 0.93 & $0.77\pmm{+0.2}{-0.4}$ & 0.54 \cr
 0.1~ & 0.3 & $0.33\pmm{+0.3}{-0.8}$ & 0.98 & $0.81\pmm{+0.2}{-0.4}$ & 0.62 \cr
 0.1~ & 0.5 & $0.35\pmm{+0.4}{-0.9}$ & 1.01 & $0.88\pmm{+0.3}{-0.5}$ & 0.69 \cr
 0.1~ & 1.0 & $0.41\pmm{+0.5}{-1.0}$ & 1.08 & $1.06\pmm{+0.4}{-0.6}$ & 0.82 \cr
 0.1~ & 1.5 & $0.48\pmm{+0.6}{-1.2}$ & 1.13 & $1.30\pmm{+0.5}{-0.8}$ & 0.90 \cr
\tablerule
 0.15 & 0.0 & $0.34\pmm{+0.3}{-0.8}$ & 1.06 & $0.81\pmm{+0.2}{-0.4}$ & 0.66 \cr
 0.15 & 0.3 & $0.35\pmm{+0.3}{-0.8}$ & 1.08 & $0.84\pmm{+0.2}{-0.5}$ & 0.71 \cr
 0.15 & 0.5 & $0.37\pmm{+0.4}{-0.9}$ & 1.12 & $0.90\pmm{+0.3}{-0.5}$ & 0.78 \cr
 0.15 & 1.0 & $0.43\pmm{+0.4}{-1.0}$ & 1.19 & $1.08\pmm{+0.4}{-0.6}$ & 0.91 \cr
 0.15 & 1.5 & $0.51\pmm{+0.6}{-0.2}$ & 1.23 & $1.32\pmm{+0.5}{-0.8}$ & 1.00 \cr
\spl
\endinsert
\vfill\dosupereject

\baselineskip=20pt
\bye